\documentclass[a4paper,prd]{revtex4-1}

\usepackage{amsmath}
\usepackage{amssymb}
\usepackage{graphicx}

\newcommand{\nnu}[2]{\ensuremath{n_{\text{#1}}^{#2}}}
\newcommand{\nnd}[2]{\ensuremath{n^{\text{#1}}_{#2}}}
\newcommand{\dnnu}[2]{\ensuremath{\delta n_{\text{#1}}^{#2}}}

\newcommand{\tnnu}[3]{\ensuremath{\tilde{n}_{\text{#1}, [{#3}]}^{#2}}}

\newcommand{\NNu}[3]{\ensuremath{N_{\text{#1}, [{#3}]}^{#2}}}
\newcommand{\dhnnu}[3]{\ensuremath{\delta \hat{n}_{\text{#1}, [{#3}]}^{#2}}}
\newcommand{\nnsq}[1]{\ensuremath{n_{\text{#1}}^{2}}}
\newcommand{\nnn}[1]{\ensuremath{n_{\text{#1}}}}

\newcommand{\mmd}[2]{\ensuremath{\mu^{\text{#1}}_{#2}}}

\newcommand{\dmmd}[2]{\ensuremath{\delta \mu^{\text{#1}}_{#2}}}

\newcommand{\ffd}[2]{\ensuremath{f^{\text{#1}}_{#2}}}

\newcommand{\M}{{\cal M}}
\newcommand{\X}{\ensuremath{\text{X}}}
\newcommand{\Y}{\ensuremath{\text{Y}}}
\newcommand{\Z}{\ensuremath{\text{Z}}}
\newcommand{\LL}{\ensuremath{\Lambda}}

\newcommand{\pda}[2]{\frac{\partial {#1}}{\partial {#2}}}

\newcommand{\tj}{\text{i}}

\begin{document}

\title{The nonlinear development of the relativistic two-stream instability}

\author{I.~Hawke$^1$, G.~L.~Comer$^2$, N.~Andersson$^1$}

\affiliation{$^1$Mathematical Sciences, University of Southampton, Southampton, SO17 1BJ, UK}
\affiliation{$^2$Department of Physics \& Center for Fluids at All Scales, Saint Louis University, St.~Louis, MO, 63156-0907, USA}

\begin{abstract}
  The two-stream instability has been mooted as an explanation for a range of astrophysical applications from GRBs and pulsar glitches to cosmology. Using the first nonlinear numerical simulations of relativistic multi-species hydrodynamics we show that the onset and initial growth of the instability is very well described by linear perturbation theory. In the later stages the linear and nonlinear description match only qualitatively, and the instability does not saturate even in the nonlinear case by purely ideal hydrodynamic effects.
\end{abstract}

\maketitle

\section{Introduction}
\label{sec:intro}

The two-stream instability occurs generically when two coupled, inter-penetrating fluids have a sufficiently large relative velocity. The expectation is that the instability arises when a perturbation appears to move in different directions with respect to each fluid. Originally studied in magnetized plasmas modeled by the collisionless Vlasov-Maxwell equations~\cite{Farley1963,Buneman1963}, where a detailed understanding of the full mode spectrum exists (see e.g.~\cite{Bret2010}), it is only recently (see~\cite{Samuelsson2009}) that it has been studied in the context of pure relativistic hydrodynamics. The well-understood plasma instability has astrophysical applications in GRBs (see e.g.~\cite{Bret2010}) and the pulsar emission mechanism (\cite{Nanobashvili2011,Sturrock2012}), whilst the hydrodynamic instability has been suggested as a key mechanism behind pulsar glitches (\cite{Andersson2003,Andersson2004}) and is of relevance in the cosmological context~\cite{Comer2012}.

Most existing explorations of the two-stream instability have focused on its development in the linear regime, as -- in astrophysically interesting applications such as pulsar glitches -- the instability is expected to saturate at relatively low amplitude. However, the mechanism by which the instability saturates is not clear. The obvious possibility is that non-ideal effects such as shear viscosity will stop the growth. However, given the difficulties in constructing a consistent and stable relativistic non-ideal theory this avenue can only be pursued phenomenologically. An alternative possibility is that the instability saturates due to nonlinear effects within the framework of the ideal theory, possibly via the generation of internal shocks converting the unstable modes to heat. A final possibility that is unlikely to be generically applicable is that the fluids may be dynamically driven out of the instability window, perhaps by external forces or non-local effects, as happens in the cosmological case (\cite{Comer2012}).

In this paper we will study the nonlinear development of the relativistic hydrodynamic two-stream instability in simplified cases using numerical simulations. For reasons detailed later we are not able to consider shock propagation. However, the simulations can investigate if shock formation is a possibility, or whether the high frequency oscillations dominate, at which point we would expect non-ideal effects to be important. Our results suggest that when the instability is triggered by small perturbations the two-stream instability grows until the solution is dominated by high frequency oscillations. By comparing with linearized time-domain solutions we see that the nonlinear coupling has only a small effect, and is insufficient to saturate the instability.

\section{The system}
\label{sec:system}

We consider the system of relativistic multiple fluids as introduced by Carter in~\cite{Carter1989} and detailed in the review of Andersson and Comer~\cite{Andersson2007}. Here the notation largely follows~\cite{Andersson2007}, assumes the existence of a spacetime metric $g_{ab}$ of signature $-+++$, uses Roman letters from the start of the alphabet -- $a, b, c, \dots$ -- as 4-spacetime indices, and from the middle -- $i, j, k, \dots$ -- as 3-space indices in the $3+1$ split. The characters $\X, \Y, \Z$ will be used as labels indicating the different fluids, or different species, which will not be implicitly summed over except where explicitly stated. Units where the speed of light $c = 1$ are used throughout.

\subsection{General form}
\label{sec:system_general}

The basic fluid quantities are the number density currents \nnu{\X}{a} where $\X$ is a species label. Here we do not consider reactions or particle creation, meaning that the number densities are conserved. Hence the currents obey the continuity equations
\begin{equation}
  \label{eq:continuity}
  \nabla_a \nnu{\X}{a} = 0.
\end{equation}
The system is closed using the master function \LL. The equations of motion -- the Euler equations -- follow by minimizing the action defined using this master function. The master function is defined in terms of all possible scalar invariants $\nnsq{\X\Y} = -g_{ab} \nnu{\X}{a} \nnu{\Y}{b}$. The shorthand notation $\nnsq{\X\X} \equiv \nnsq{\X}$ is often used. The Euler equations are written in terms of the conjugate momenta $\mmd{\X}{a}$, defined by
\begin{equation}
  \label{eq:conjugate_momenta}
  \mmd{\X}{a} = \pda{\LL}{\nnu{\X}{a}} = -2 \pda{\LL}{\nnsq{\X}} \nnd{\X}{a} - \sum_{\Z \neq \X} \pda{\LL}{\nnsq{\X\Z}} \nnd{\Z}{a}.
\end{equation}
Using these definitions the Euler equations follow as
\begin{equation}
  \label{eq:Euler_eqns}
  \ffd{\X}{b} \equiv 2 \nnu{\X}{a} \nabla_{[a} \mmd{\X}{b]} = 0.   
\end{equation}

\subsection{$3+1$ decomposition}
\label{sec:system_3plus1}

As a test-bed we will consider flat spacetimes in standard plane symmetric Cartesian coordinates. Using $(t, i)$ to represent the time and spatial coordinates, the continuity equation~\eqref{eq:continuity} becomes
\begin{subequations}
  \label{eq:eom_3plus1}
  \begin{align}
    \label{eq:continuity_3plus1}
    \partial_t \nnu{\X}{t} + \partial_i \nnu{\X}{i} &= 0 \\
    \intertext{and the Euler equation~\eqref{eq:Euler_eqns} can be written as}
    \label{eq:Euler_eqns_3plus1}
    \partial_t \mmd{\X}{i} - \partial_i \mmd{\X}{t} - 2 \frac{\nnu{\X}{j}}{\nnu{\X}{t}} \partial_{[i} \mmd{\X}{j]} &= 0.
  \end{align}
\end{subequations}

\subsection{Numerical implementation}
\label{sec:system_numerics}

The nonlinear numerical simulations solve the equations of motion~\eqref{eq:eom_3plus1}. Although these describe conservative ideal (multi-species) hydrodynamics, they are not written in balance law form. Therefore discontinuous ``solutions'' will not necessarily be the correct (entropy satisfying) solution, irrespective of the numerical techniques employed -- the derivation of a suitable entropy satisfying balance law form is required future work before multifluid systems with shocks can be studied. For simplicity we will therefore use a basic finite difference discretization of the equations of motion~\eqref{eq:eom_3plus1} and note that the numerical solutions should not be trusted in two regimes: firstly, when the spatial gradients are sufficiently large to appear discontinuous on the grid, and secondly when the solution is dominated by high frequency (relative to the numerical grid) components.

The implementation used here relies on the Method of Lines where the time integration uses either the standard third order strict stability preserving (SSP) Runge-Kutta method, or the standard fourth order non-SSP Runge-Kutta method. The spatial discretization uses central differencing of either second or fourth order. We have tested whether the addition of Kreiss-Oliger dissipation (of either third or fifth order respectively) makes any difference, and it does not. In all simulations we set the timestep by imposing a CFL constant $\Delta t / \Delta x$ of $0.25$, and have checked that the results are insensitive to the precise value.

The numerical solution of Eq.~\eqref{eq:eom_3plus1} gives solutions for $\nnu{\X}{t}$ and $\mmd{\X}{i}$. This is sufficient to describe the system completely, but in order to numerically compute the terms for the next update it is necessary to construct the spatial components of the number density flux, $\nnu{\X}{i}$. To do this we note that the definition of the conjugate momenta, Eq.~\eqref{eq:conjugate_momenta}, can be seen as a system of nonlinear algebraic equations for the $\nnsq{\X}$. Solving this nonlinear algebraic system gives the correct value for the scalars $\nnsq{\X}$, from which we can compute the master function $\LL$ and its derivatives. This allows Eq.~\eqref{eq:conjugate_momenta} to be viewed as a linear system for the spatial components $\nnu{\X}{i}$, as required.

\section{Linearized solution}
\label{sec:linear_solution}

To study the time domain behaviour of the multifluid system, and as a comparison test for the nonlinear simulations, we construct linearized solutions of two coupled fluids on a $2 L$ periodic domain $[-L, L]$ in one spatial dimension. The restrictions on the number of fluids and spatial dimensions can be dropped in the following analysis at the cost of substantially complicating the explicit calculation of the final solution.

\subsection{Linearized system}
\label{sec:linear_solution_system}

Start by restricting the equations of motion~\eqref{eq:eom_3plus1} to one spatial dimension $x$, which after linearizing gives the system
\begin{subequations}
  \label{eq:linear_system_1}
  \begin{align}
    \partial_t \dnnu{X}{t} + \partial_x \dnnu{X}{x} & = 0, \\
    \partial_t \dmmd{X}{x} - \partial_x \dmmd{X}{t} & = 0.
  \end{align}
\end{subequations}
Note that, from the definition of the conjugate momenta given in Eq.~\eqref{eq:conjugate_momenta}, we have (see~\cite{Andersson2007})
\begin{equation}
  \label{eq:linear_system_delta_mu}
  \delta\mmd{\X}{a} = \M^{\X\Z}_{ab} \, \dnnu{\Z}{b}
\end{equation}
where here, and in the future, $\Z$ is an abstract species index to be summed over, and where the matrices can be given explicitly (e.g.~\cite[section 11.3]{Andersson2007}).

We note that the linearized system can be written as
\begin{subequations}
  \label{eq:linear_system_2}
  \begin{align}
    \partial_t \dnnu{X}{t} + \partial_x \dnnu{X}{x} & = 0, \\
    \M^{\X\Z}_{xx} \partial_t \dnnu{Z}{x} - \left( \M^{\X\Z}_{xt} + \M^{\X\Z}_{tx} \right) \partial_x \dnnu{Z}{x} - \M^{\X\Z}_{tt} \partial_x \dnnu{Z}{t} & = 0.
  \end{align}
\end{subequations}
We will not use this form in the linearized solution, but will use Eq.~\eqref{eq:linear_system_2} to construct numerical solutions, using the techniques of Sec.~\ref{sec:system_numerics}, for direct comparison with the nonlinear results.

\subsection{Transformed system}
\label{sec:linear_solution_transform}

We now use a Fourier-Laplace analysis, firstly taking the discrete Fourier transforms by writing
\begin{equation}
  \label{eq:linear_solution_fourier}
  \dnnu{\X}{a} = \sum_{k=0}^{k_{\text{max}}} \tnnu{\X}{a}{k} \exp[ \tj \omega_k x]
\end{equation}
where $\omega_k = \pi k / L$. Then the Laplace transform
\begin{equation}
  \label{eq:linear_solution_laplace_1}
  \NNu{\X}{a}{k} = {\cal L} \left[ \tnnu{\X}{a}{k} \right]
\end{equation}
leads to the fully transformed equations
\begin{subequations}
  \label{eq:linear_solution_transformed_system}
  \begin{align}
    \label{eq:linear_solution_transformed_system_t}
    s \NNu{\X}{t}{k} + \tj \omega_k \NNu{\X}{x}{k} & = \tnnu{\X}{t}{k}(t = 0), \\
    \label{eq:linear_solution_transformed_system_x}
    s \left[ \M^{\X\Z}_{a x} \NNu{\Z}{a}{k}\right] - \tj \omega_k \left[ \M^{\X\Z}_{a t} \NNu{\Z}{a}{k} \right] & = \M^{\X\Z}_{a x} \tnnu{\Z}{a}{k}(t = 0) .
  \end{align}
\end{subequations}

\subsection{Solution}
\label{sec:linear_solution_solution}

Combining the results from Eq.~\eqref{eq:linear_solution_transformed_system}, the solution for $\NNu{\X}{x}{k}$ follows from the linear system
\begin{equation}
  \label{eq:linear_solution_formal}
  A^{\X\Z} \NNu{\Z}{x}{k} = \hat{V}^{\X} 
\end{equation}
where the matrix $A$ is given by
\begin{equation}
  \label{eq:linear_solution_matrix}
  A^{\X\Z} = s^2 \M^{\X\Z}_{xx} - s \tj \omega_k \left( \M^{\X\Z}_{xt} + \M^{\X\Z}_{xt} \right) + \omega_k^2 \M^{\X\Z}_{tt}
\end{equation}
and the vector $\hat{V}$ by
\begin{equation}
  \label{eq:linear_solution_vector}
  \hat{V}^{\X} = s \M^{\X\Z}_{xx} \dhnnu{\Z}{x}{k} + \tj \omega_k \M^{\X\Z}_{tt} \dhnnu{\Z}{x}{m}. 
\end{equation}
The solution for $\NNu{\X}{t}{k}$ follows from Eq.~\eqref{eq:linear_solution_transformed_system_t}. Given these solutions we can invert the Laplace and Fourier transforms to construct $\dnnu{\X}{a}$ at any future time.

\subsection{The instability}
\label{sec:linear_solution_instability}

We note that the qualitative change from a stable two-fluid system to an unstable one can easily be seen with this linearized solution. In the formal solution given by Eq.~\eqref{eq:linear_solution_formal} the inverse of the matrix $A^{\X\Z}$ can be written
\begin{equation}
  \label{eq:linear_solution_instability_a_inv}
  \left( A^{\X\Z} \right)^{-1} = \frac{\text{adj} A^{\X\Z}}{\alpha \prod_{i=1}^{4} (s - \tj \omega_k r_i)}
\end{equation}
where the denominator, a quartic polynomial in $s$, is the determinant of $A^{\X\Z}$. This form is chosen so that the roots $r_i$ are independent of the frequency $\omega_k$. On inverting the Laplace transform to construct the solution for $\dnnu{\X}{a}$ we find that the time dependent behaviour is encoded in exponentials of the form $\exp(\tj \omega_k r_i t)$. Therefore, the resulting linearized solution is stable only if the roots $r_i$ have vanishing imaginary part and, when unstable, the growth rate is linearly related to the frequency $\omega_k$ as expected. This precisely mirrors previous calculations (e.g.~\cite{Samuelsson2009}) where the stability was usually found more straightforwardly in terms of the dispersion relation.

\section{Results}
\label{sec:results}

In what follows we shall always use an equation of state inspired by~\cite{Comer2012} and \cite{Prix2005} and encompassing the key expected behaviour. First note (\cite{Prix2005}) that the Lorentz factor of one fluid in the frame of the other is
\begin{equation}
  \label{eq:relative_lorentz}
  W_{12} = - \frac{\nnu{a}{1} \nnu{b}{2}}{\nnn{1} \nnn{2}} g_{ab} = \frac{\nnn{12}^2}{\nnn{1} \nnn{2}} = \left( 1 - \Delta^2 \right)^{-1/2},
\end{equation}
implying that $\Delta$, defined by
\begin{equation}
  \label{eq:eos_delta}
  \Delta^2 = 1 - \left( \frac{ \nnn{1} \nnn{2} }{ \nnsq{12} } \right)^2,
\end{equation}
encodes the velocity difference between the two fluids.
Then we use a master function with general form 
\begin{equation}
  \label{eq:eos}
  -\LL \left( \nnsq{1}, \nnsq{2}, \nnsq{12} \right) = \sum_{\X = 1}^{2} \left( m_{\X} \nnn{\X} + \kappa_{\X} \nnn{\X}^{\gamma_{\X}} \right) + \kappa_{12} \nnn{1}^{\sigma_1} \nnn{2}^{\sigma_2} + \kappa_{\Delta} \Delta.
\end{equation}
We would approximately expect the $\kappa_{\Delta}$ term to encode the \emph{entrainment} effects, and the $\kappa_{12}$ term to encode the \emph{chemical coupling}.

\begin{table}
  \centering
  \begin{tabular}{c|c c c c c c c c c c}
    & $m_1$ & $m_2$ & $\kappa_1$ & $\kappa_2$ & $\gamma_1$ & $\gamma_2$ & $\kappa_{12}$ & $\sigma_1$ & $\sigma_2$ & $\kappa_{\Delta}$ \\ \hline
    Entrainment & 1 & 1 & $\tfrac{1}{2}$ & $\tfrac{1}{2}$ & $\tfrac{8}{5}$ & $\tfrac{9}{5}$ & 0 & N/A & N/A & $p$ \\
    Chemical coupling & 0 & 0 & $2^{-1/2}$ & 1 & 1 & $\tfrac{4}{3}$ & $p$ & 1.1 & 1.1 & 0
  \end{tabular}
  \caption{The parameters used in the simulations below, based on the master function given in Eq.~\eqref{eq:eos}. In each case one parameter, indicated with $p$, controls the coupling between the two species. In the pure entrainment case where $\kappa_{12}=0$ the values of $\sigma_{\X}$ are irrelevant.}
  \label{tab:eos_parameters}
\end{table}

\begin{figure}
  \centering
  \includegraphics[width=0.9\textwidth]{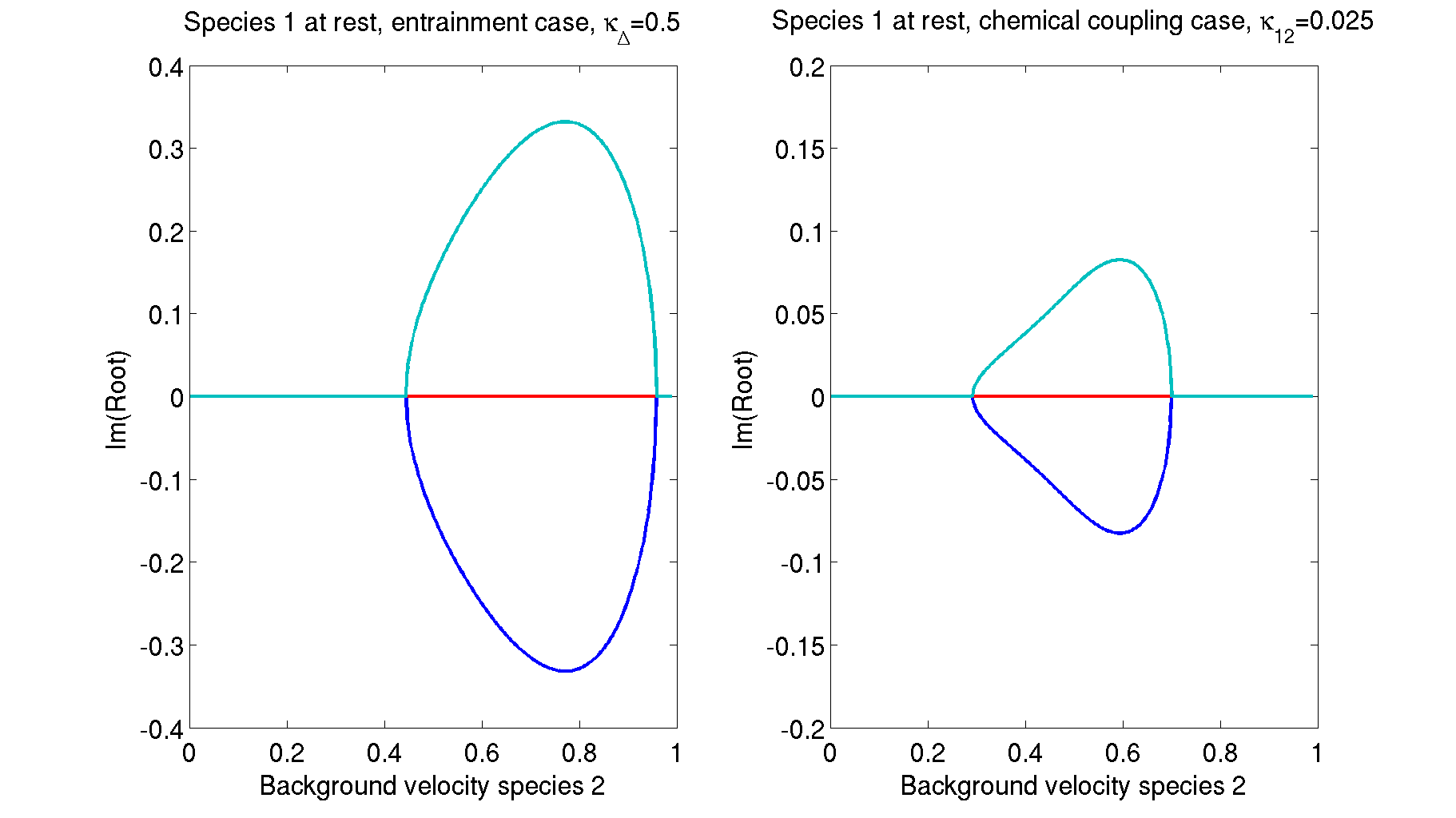}
  \caption{The two-stream instability acts when the roots of the determinant of a particular matrix (see Eq.~\eqref{eq:linear_solution_instability_a_inv} and accompanying text) have non-vanishing imaginary part. The left plot shows the entrainment case, and the right the chemical coupling case -- the full parameters are outlined in the text and Table~\ref{tab:eos_parameters}. For this particular choice there is little qualitative difference in the  instability window.}
  \label{fig:background_stability}
\end{figure}

We look at the entrainment and chemical coupling cases separately,
insofar as this separation makes sense at the nonlinear level. The
precise parameters employed are given in
Table~\ref{tab:eos_parameters}. In the pure entrainment case the
strength of the coupling is determined by the parameter
$\kappa_{\Delta}$. For the pure chemical coupling case we consider
parameters relevant to cosmology (see~\cite{Comer2012}), where the
strength of the coupling is determined by the parameter
$\kappa_{12}$. The chemical coupling case also checks that the methods
work in the limit $m_{\X} \to 0$ as
expected. Fig.~\ref{fig:background_stability} shows the instability
window for these representative cases where the background is given by
the simple choice $\nnsq{\X} = 1$.

The initial data chosen here is always written as a perturbation about a background, even in the nonlinear case. We write (in $1+1$ dimensions)
\begin{equation}
  \label{eq:results_ID}
  \nnu{\X}{a} = \nnn{\X} W_{\X} \left( -1, v_{\X}, 0, 0 \right), \qquad W_{\X} = \left( 1 - v_{\X}^2 \right)^{-1/2}
\end{equation}
where $W_{\X}$ is the Lorentz factor for the $\X$ species. Perturbations about the background to either $\nnn{\X}$ or $v_{\X}$ (or both) have been tested. In what follows we shall concentrate on perturbations in $\nnn{\X}$, as no qualitative difference between the cases has been found. In the majority of the cases we focus on sinusoidal perturbations of amplitude $\delta$, with different frequencies compatible with the size of the domain. These show clean behaviour at very low numerical resolution and are easy to study. Other perturbations (such as the Gaussian shown below) have been tested with no qualitative change in the key features, but may be less flexible (because of the periodic boundaries) and may require higher resolution (due to the formation of steep gradients without forming discontinuities).

\subsection{Convergence}
\label{sec:results_convergence}

\begin{figure}
  \centering
  \includegraphics[width=0.9\textwidth]{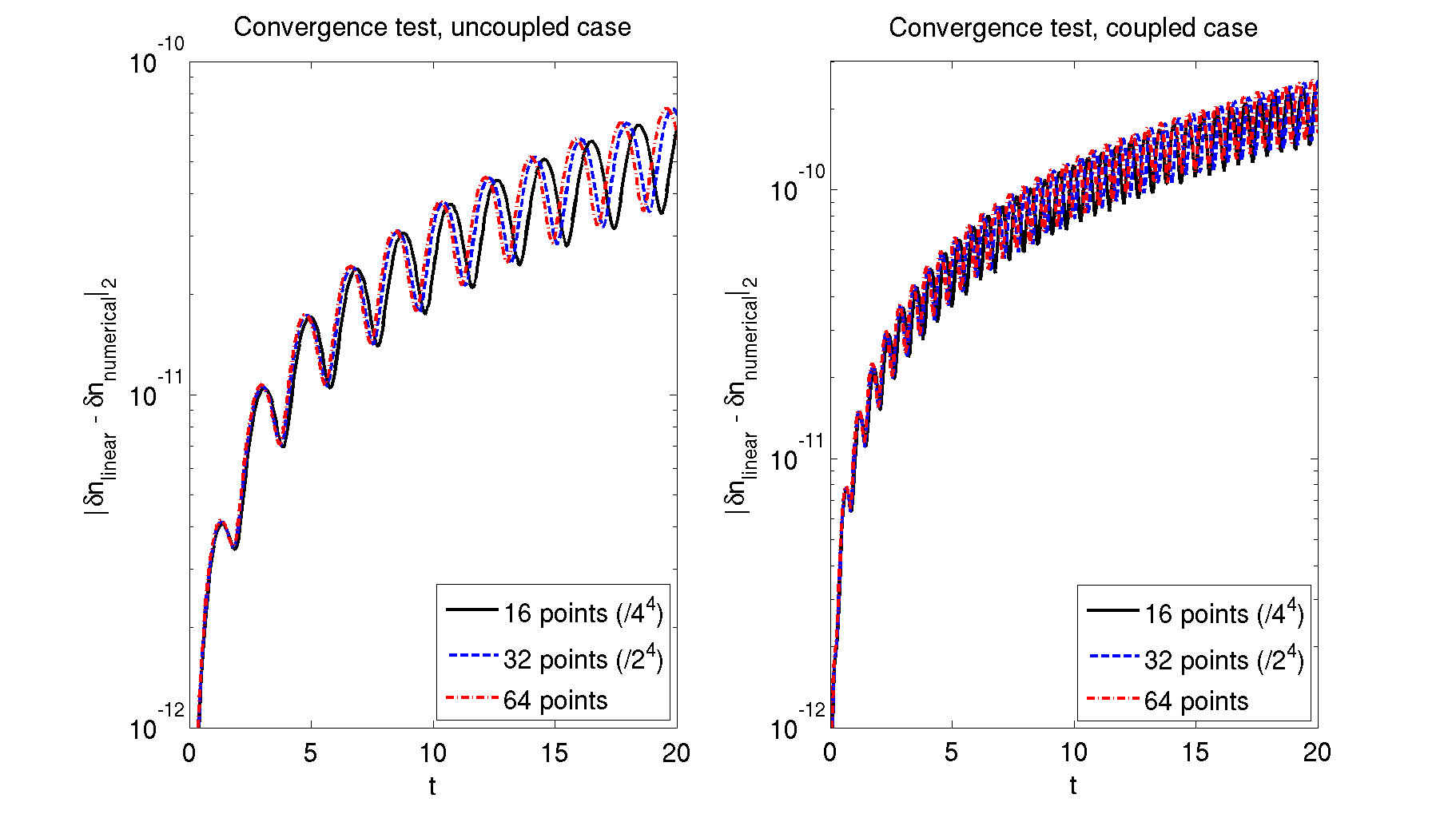}
  \caption{Convergence tests of the nonlinear code against the linearized solution. In both cases the nonlinear code is fully fourth order accurate and the curves are scaled to show the fourth order convergence. The entrainment case is shown, and in both cases the perturbation is a sine wave period $2 \pi$, amplitude $\delta = 10^{-6}$, in $\nnn{1}$. In the left panel the fluids are uncoupled, $\kappa_{\Delta} = 0$, and both are at rest in the background. In the right panel the fluids are coupled with $\kappa_{\Delta} = 0.5$ and the first fluid also has a non-trivial background velocity of $0.1$ whilst the second is at rest.}
  \label{fig:convergence1}
\end{figure}

Firstly we use the linearized solution outlined in section~\ref{sec:linear_solution} to benchmark the nonlinear code in the stable regime. This is only possible for a certain range of initial perturbation amplitudes, outside of which the convergence of the error is limited by nonlinear effects or floating point precision. In Fig.~\ref{fig:convergence1} we show simple convergence tests for a single mode perturbation in the entrainment case -- similar results are seen in the chemical coupling case. The master function parameters are as detailed above. In addition, for the uncoupled case we have $\kappa_{\Delta} = 0$ and for the coupled case we choose $\kappa_{\Delta} = 1/2$. The uncoupled case is the simplest possible -- both fluids are at rest in the background with $\nnn{\X} = 1$. In the coupled case a velocity difference is imposed by giving the first fluid a velocity of $0.1$. The perturbation is imposed only in the first fluid and is a simple sine perturbation, period $2 \pi$, amplitude $\delta = 10^{-6}$, in the number density $\nnn{1}$. Similar results are seen when the initial perturbation is a Gaussian, although considerably higher resolution is required to resolve the spatial gradients.

We note that the results converge as expected for low resolutions. At higher resolution we typically see results that are not perfectly convergent, either due to the tiny nonlinear couplings starting to dominate over the linear effects, or because the numerical error is affected by floating point precision. Where the dominant effect is the nonlinear couplings we still see the expected \emph{self}-convergence of the nonlinear code.

\begin{figure}
  \centering
  \includegraphics[width=0.49\textwidth]{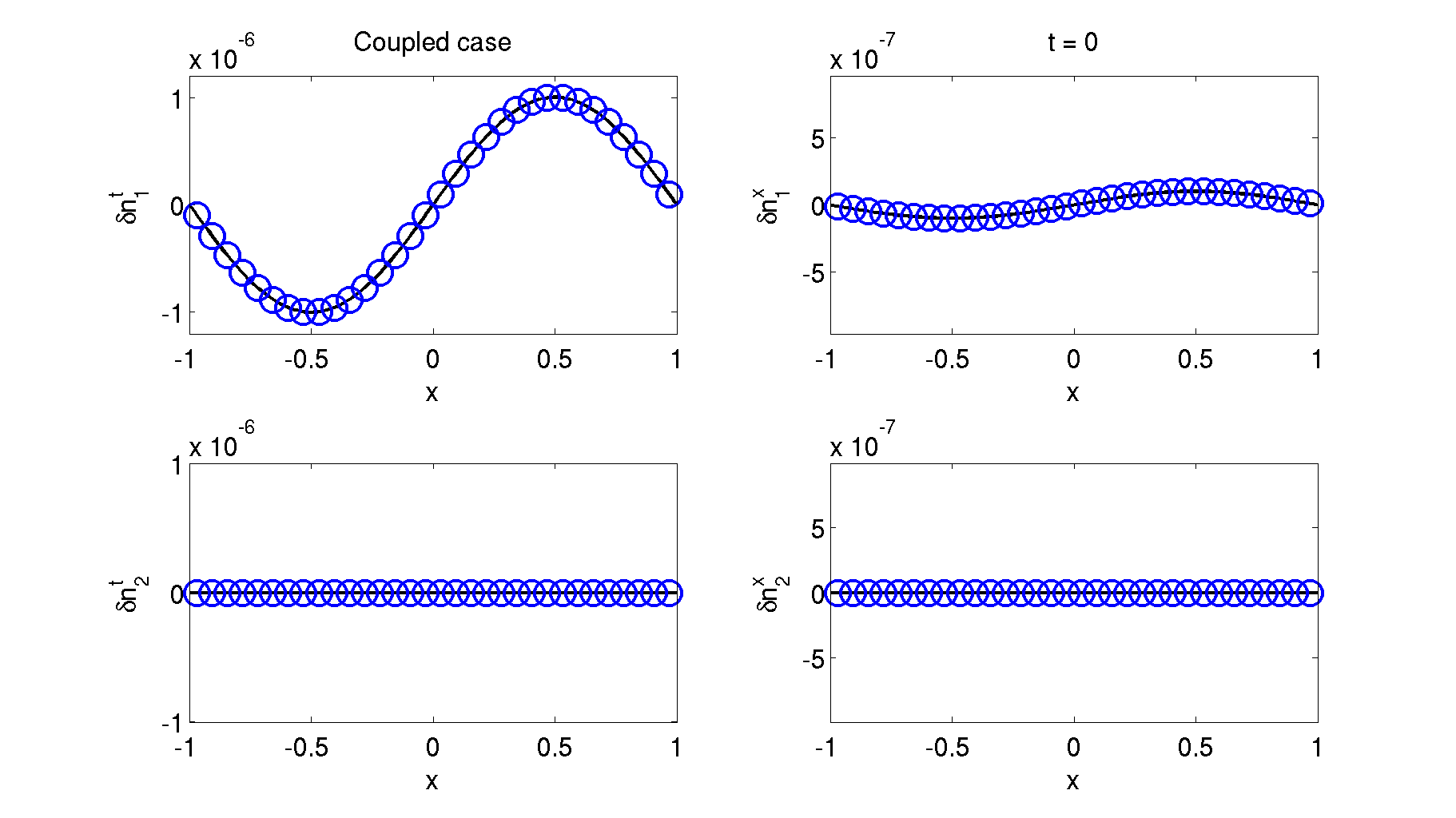}
  \includegraphics[width=0.49\textwidth]{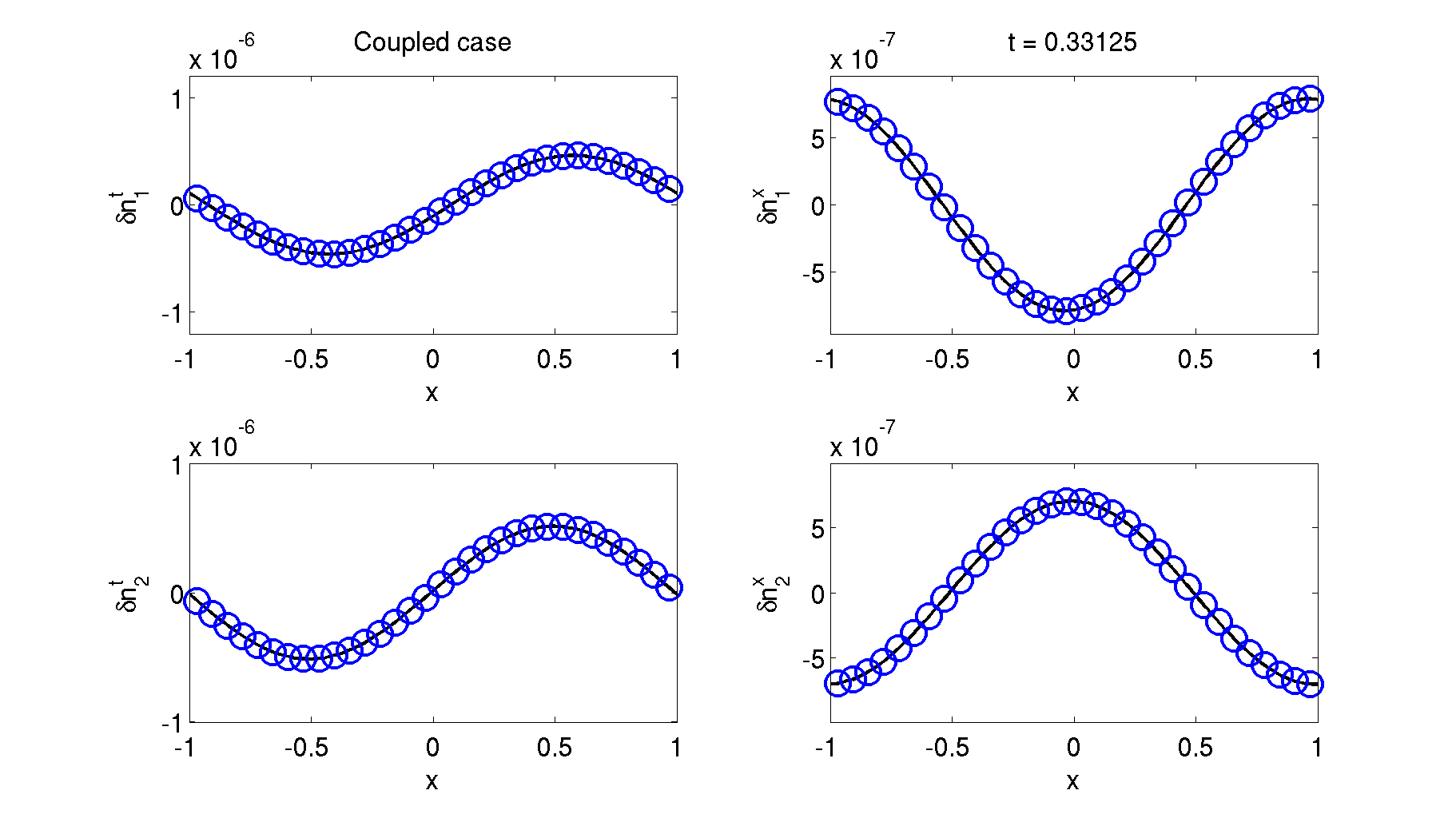}\\
  \includegraphics[width=0.49\textwidth]{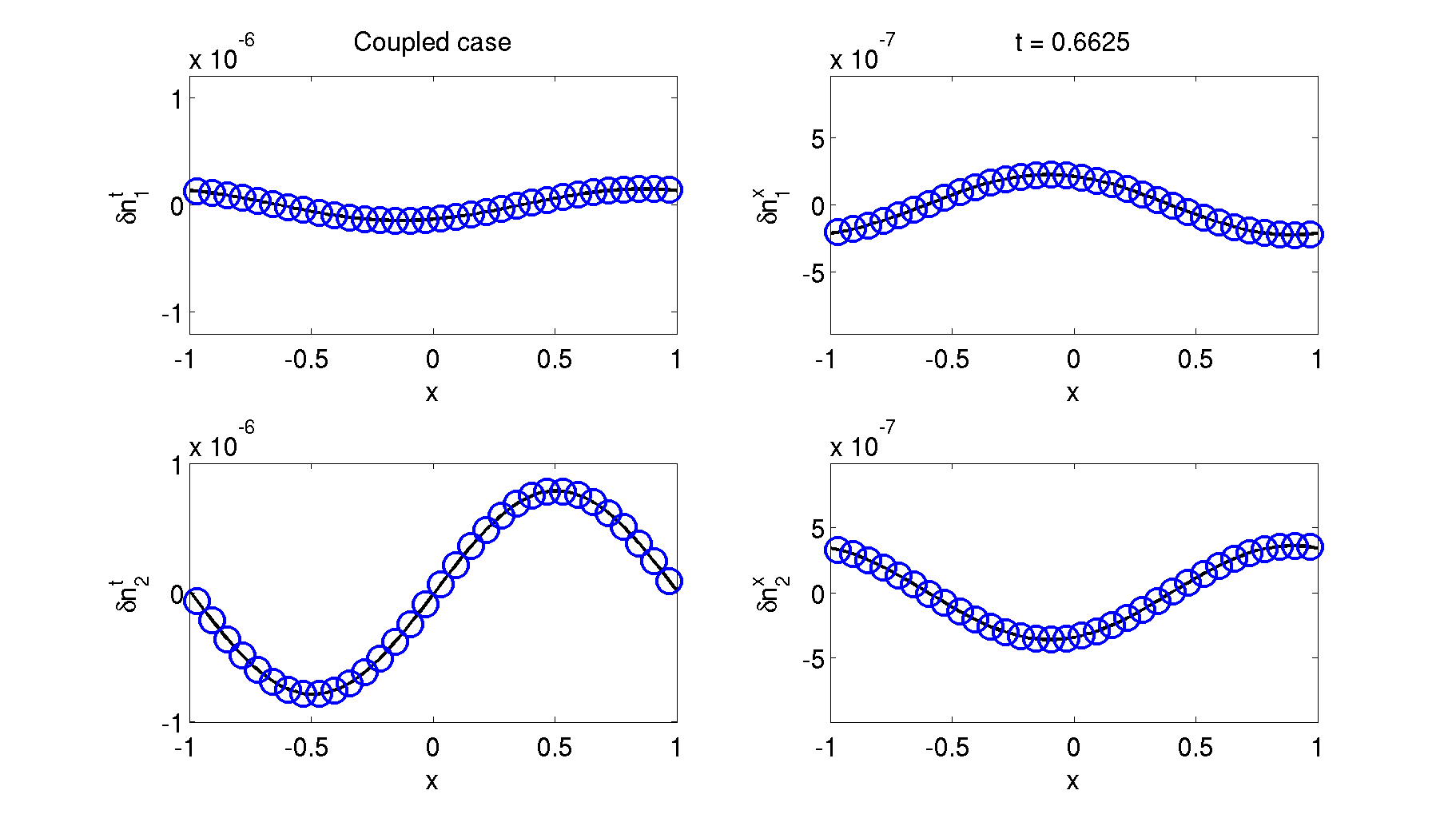}
  \includegraphics[width=0.49\textwidth]{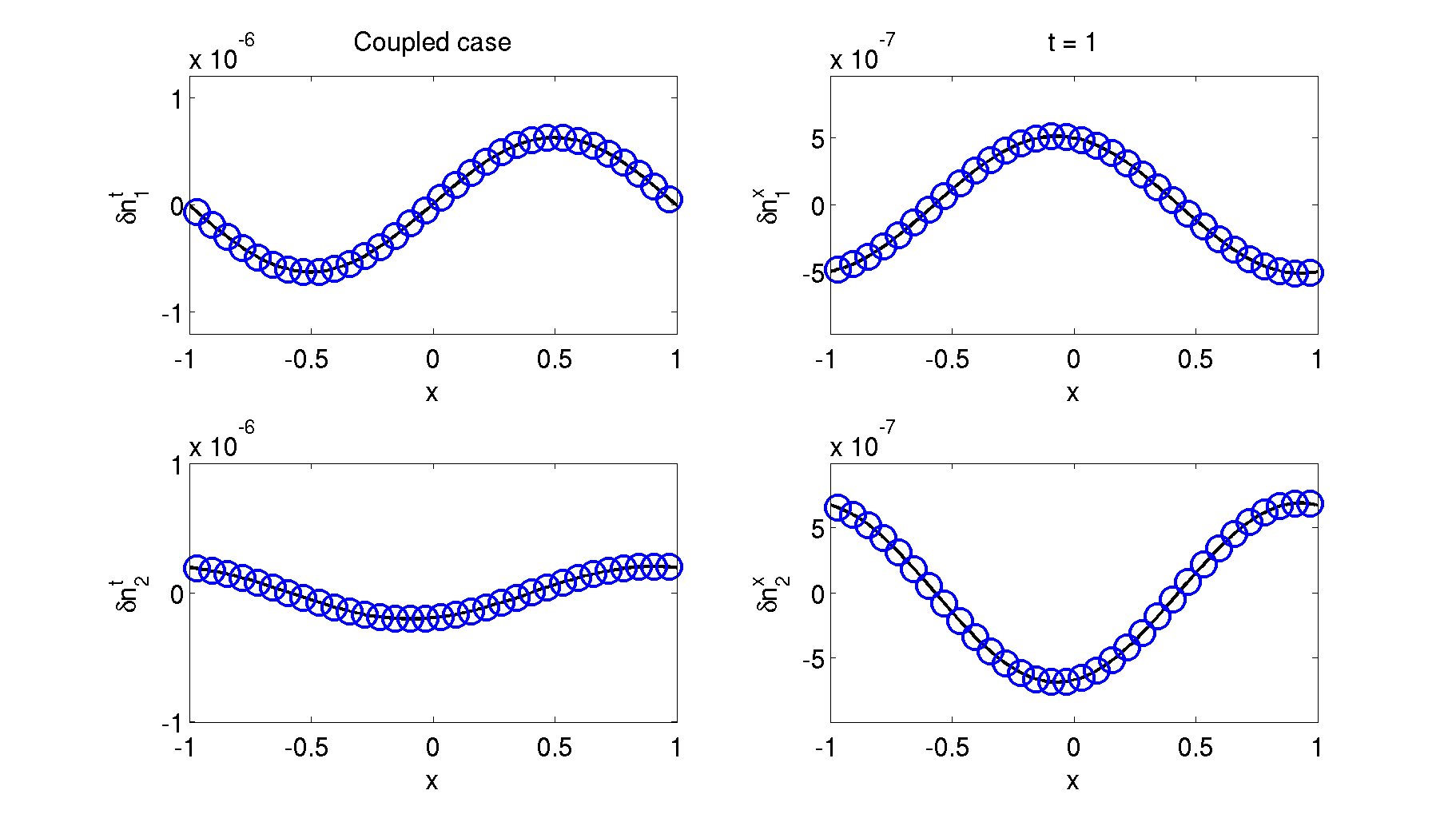}
  \caption{A direct comparison of the time evolution of the number density components in the coupled case used in the convergence plot in Fig.~\ref{fig:convergence1}. The linearized solution is given by the solid black lines, and the nonlinear numerical evolution (using only 32 points) is given by the blue circles. With these parameters we see that the fluids are fully coupled within one crossing time, but that the profiles retain the general shape given by the initial perturbation.}
  \label{fig:convergence2}
\end{figure}

A direct comparison of the components of the number densities is given in Fig.~\ref{fig:convergence2}. Over the period of approximately one crossing time we see the excellent agreement between the linearized solution and the nonlinear evolution, even at extremely low numerical resolution. The fluids rapidly couple with this choice of entrainment parameter -- within this time, which is approximately one crossing time, the amplitude of the second fluid is comparable to that of the first which contained the perturbation. The form of the perturbation is also retained through the evolution.

\subsection{Instability growth}
\label{sec:results_instability}

\begin{figure}
  \centering
  \includegraphics[width=0.49\textwidth]{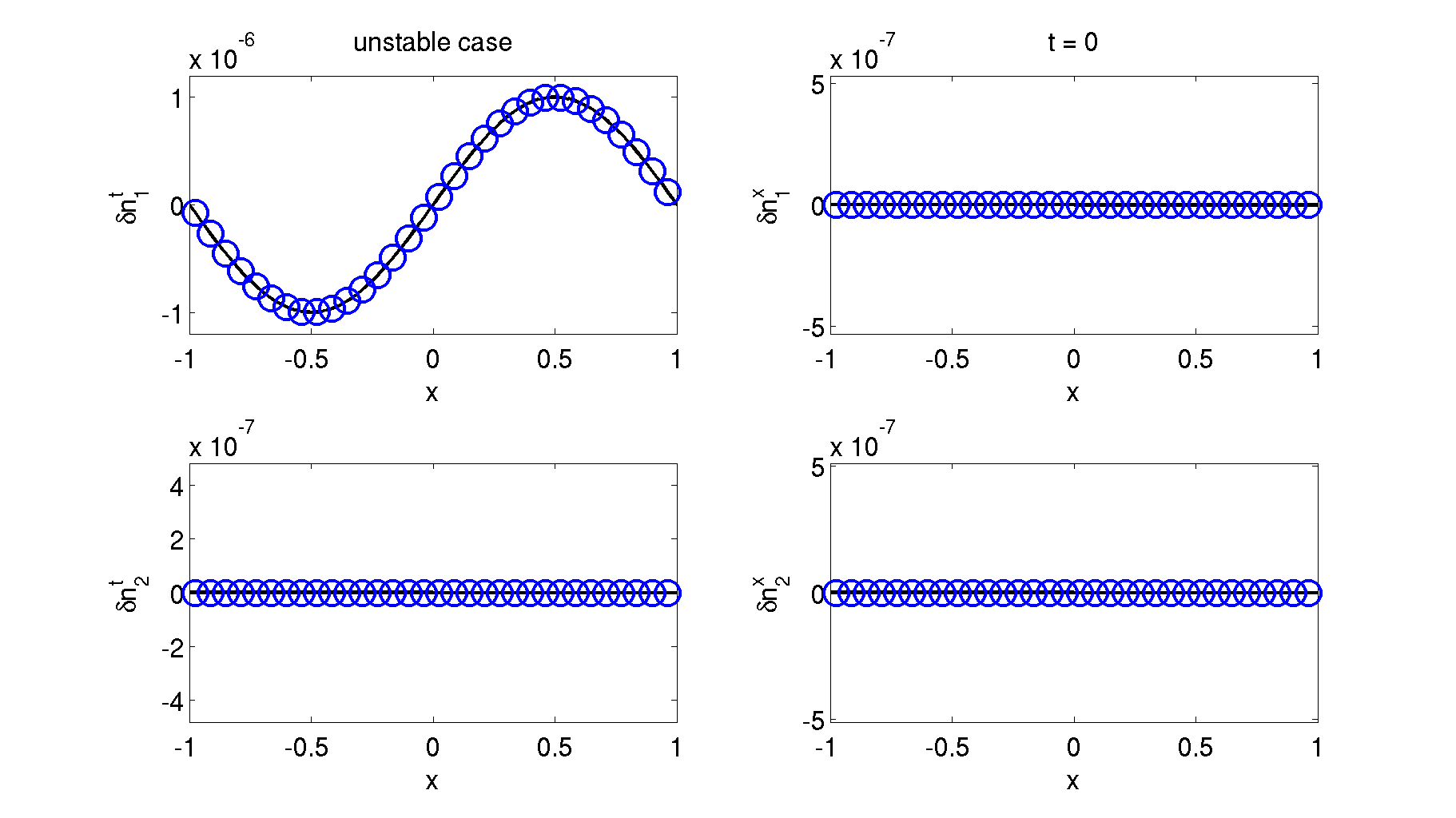}
  \includegraphics[width=0.49\textwidth]{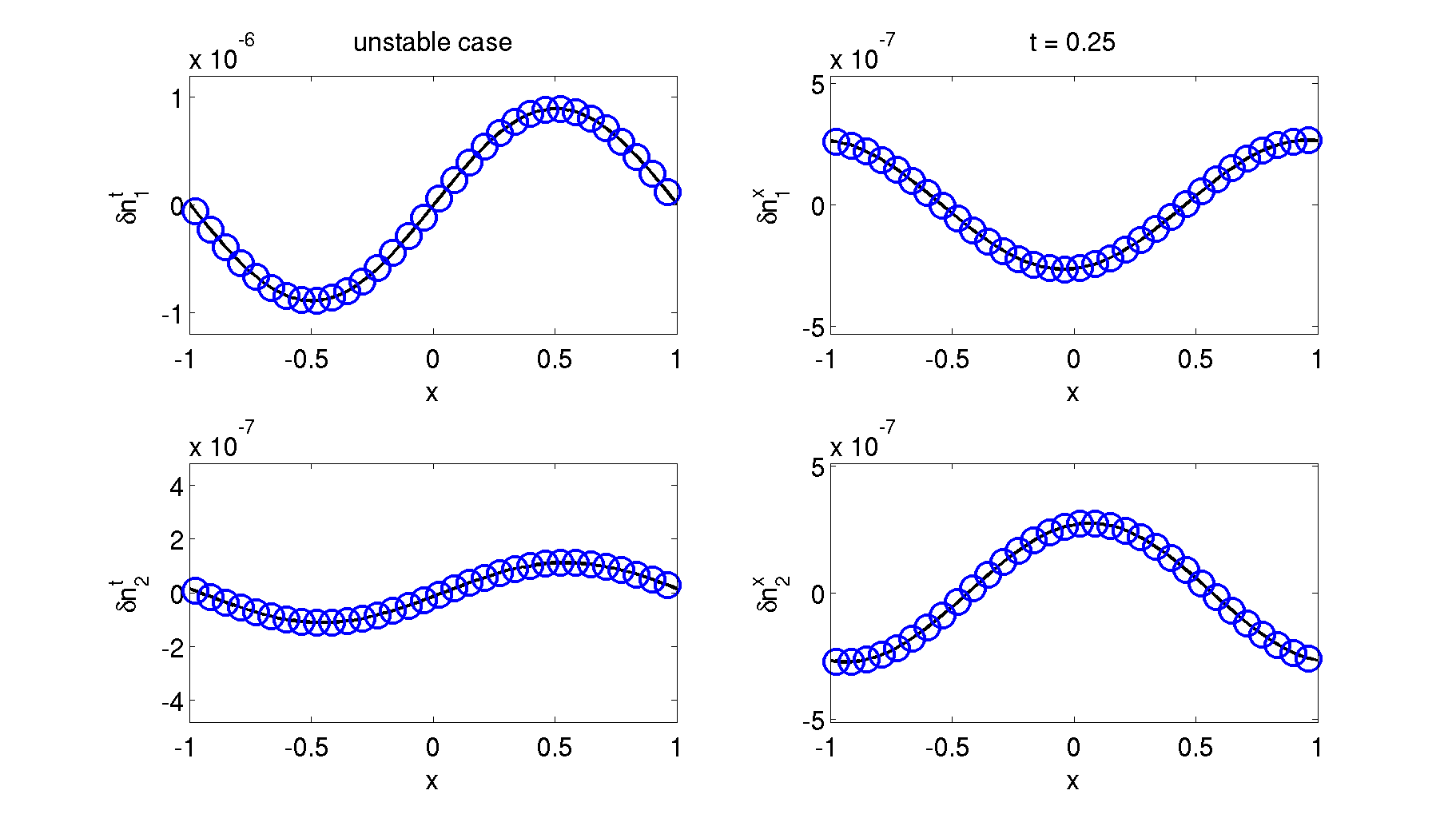}\\
  \includegraphics[width=0.49\textwidth]{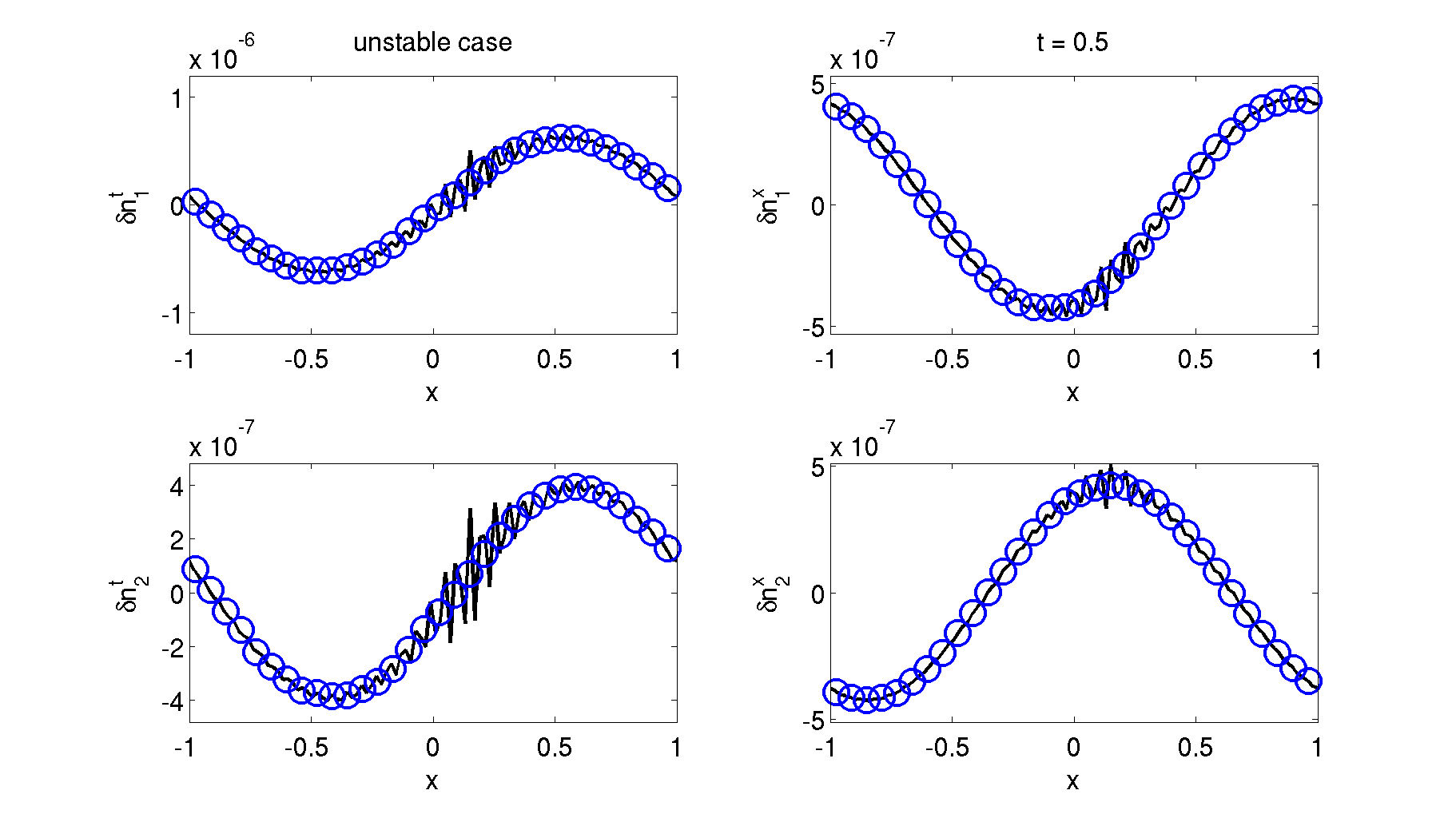}
  \includegraphics[width=0.49\textwidth]{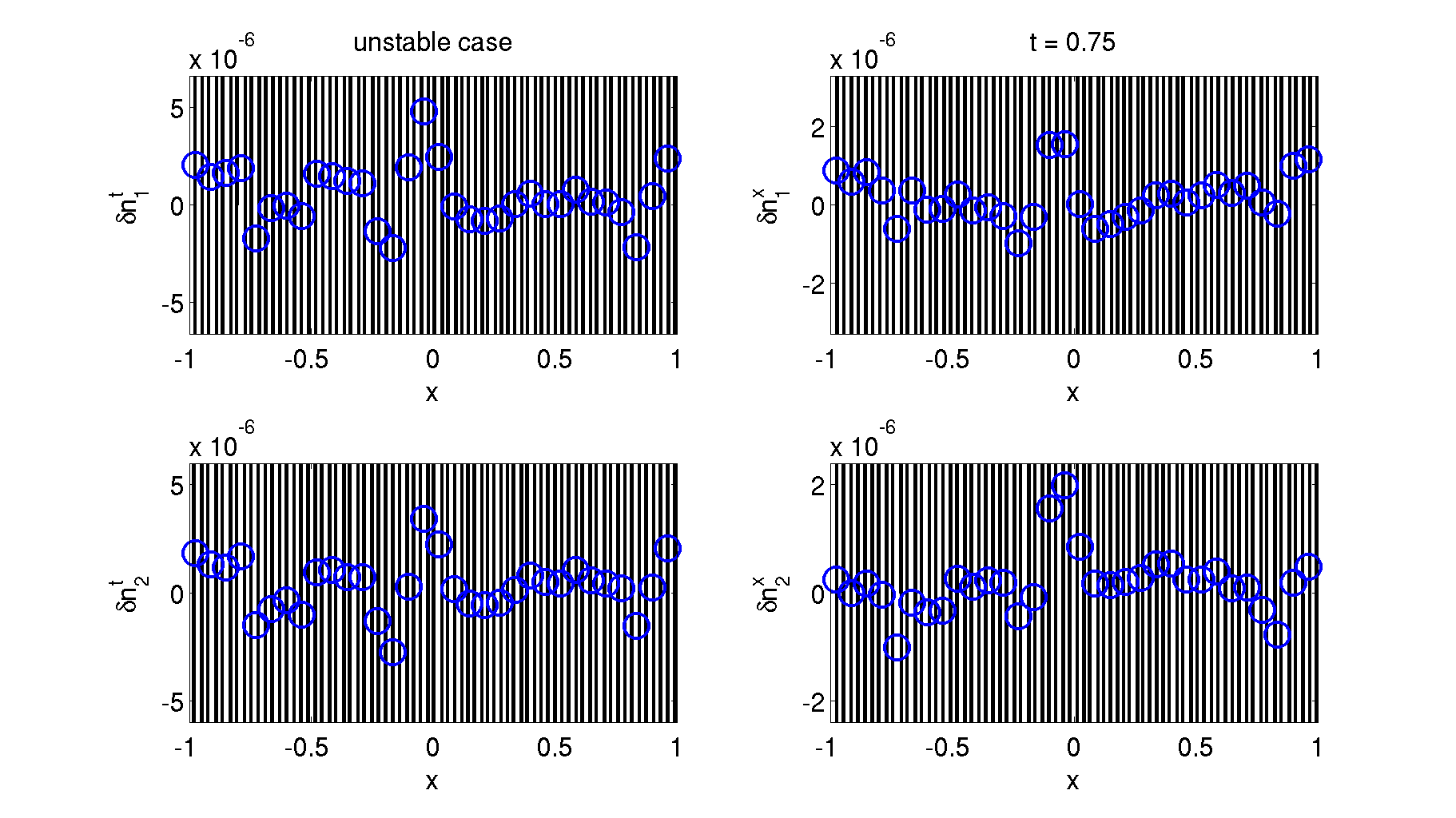}
  \caption{A direct comparison of the time evolution of the number density components in the coupled case within the instability window for the two-stream instability. The linearized solution is given by the solid black lines, and the nonlinear numerical evolution (using 128 points - every fourth shown) is given by the blue circles. This is the entrainment case with $\kappa_{\Delta} = 0.5$ and a background velocity difference of $0.6$. The instability grows rapidly in both linear and nonlinear case; to compare, the linear solution seeds all modes up to $k = 64$ with small random noise. In the early stages the behaviour closely follows the stable case shown in Fig.~\ref{fig:convergence2}. By $t = 0.5$ (bottom left panel) the instability is visible in the high-frequency oscillations for the linear solution (solid line), but is difficult to see in the nonlinear numerical solution (circles). By $t = 0.75$ (bottom right panel) the instability is clear  in both the nonlinear numerical solution and the linear solution, where the full extent of the oscillations no longer fit within the scale of the plot. The two-stream instability grows fastest in the highest frequency modes, seeded by hand in the linear case and by numerical error in the nonlinear case. The exponential growth is not spatially localized and the coupling shows the instability in all components of all species.}
  \label{fig:Instability1}
\end{figure}

\begin{figure}
  \centering
  \includegraphics[width=0.49\textwidth]{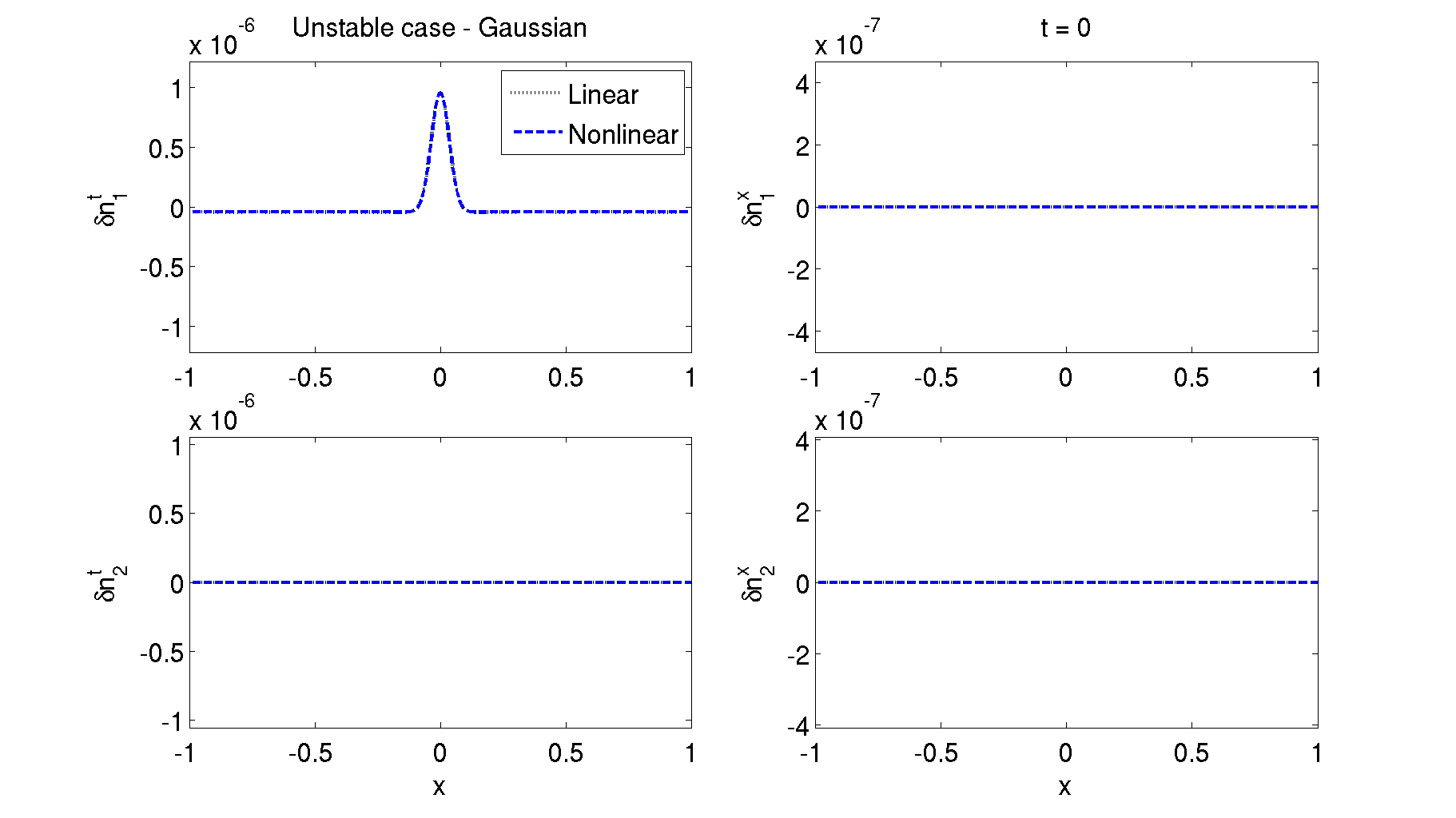}
  \includegraphics[width=0.49\textwidth]{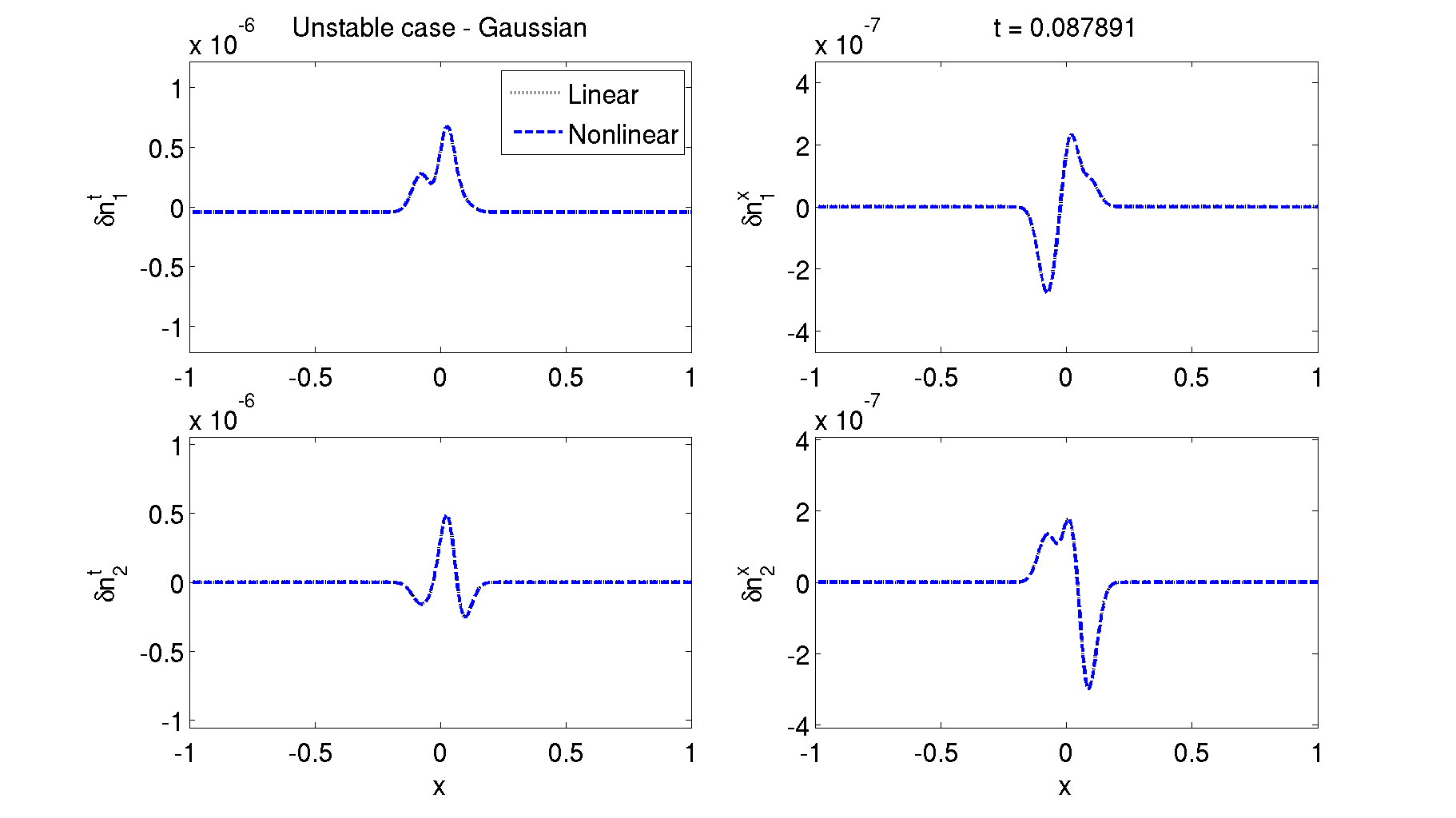}\\
  \includegraphics[width=0.49\textwidth]{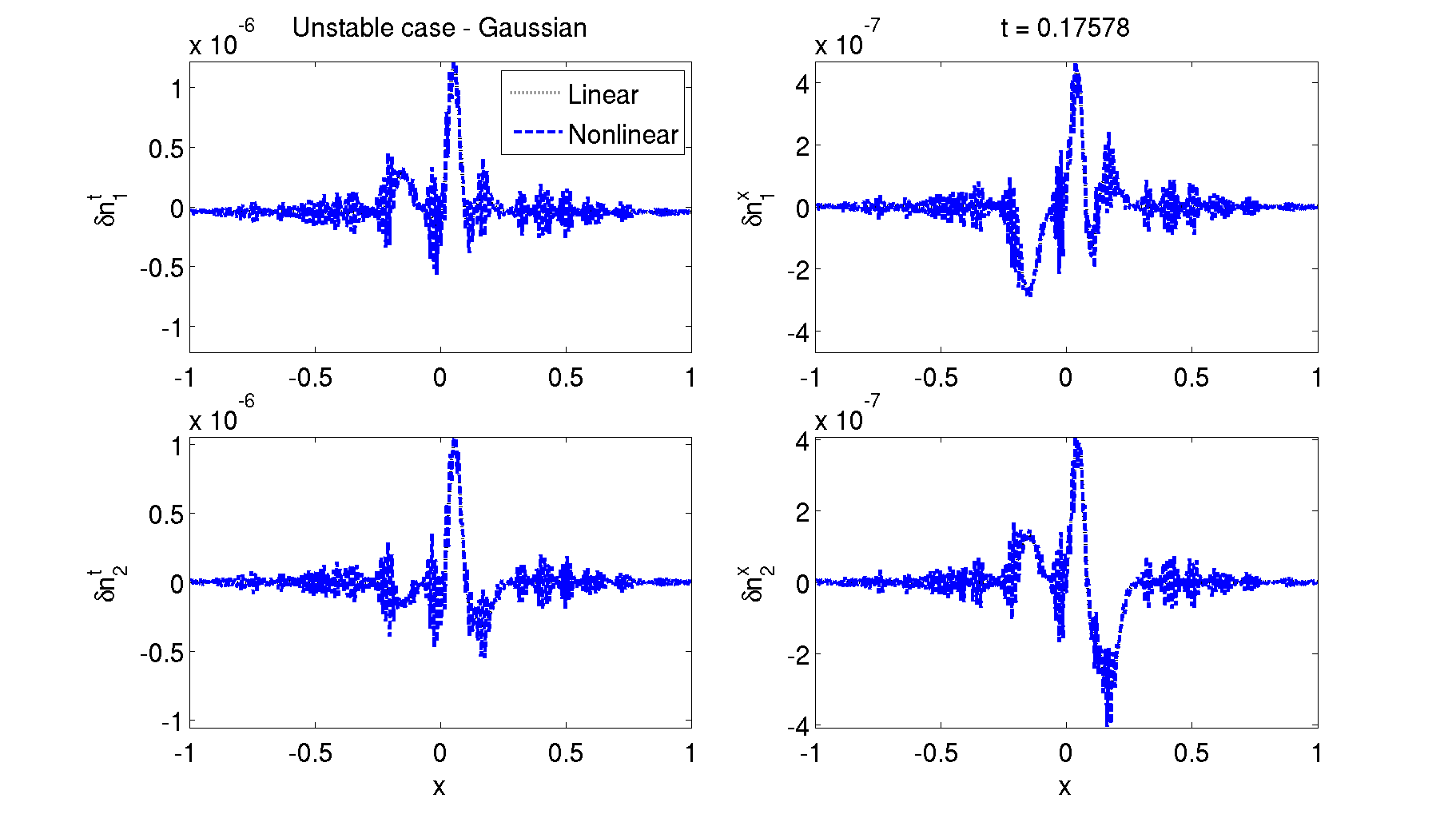}
  \includegraphics[width=0.49\textwidth]{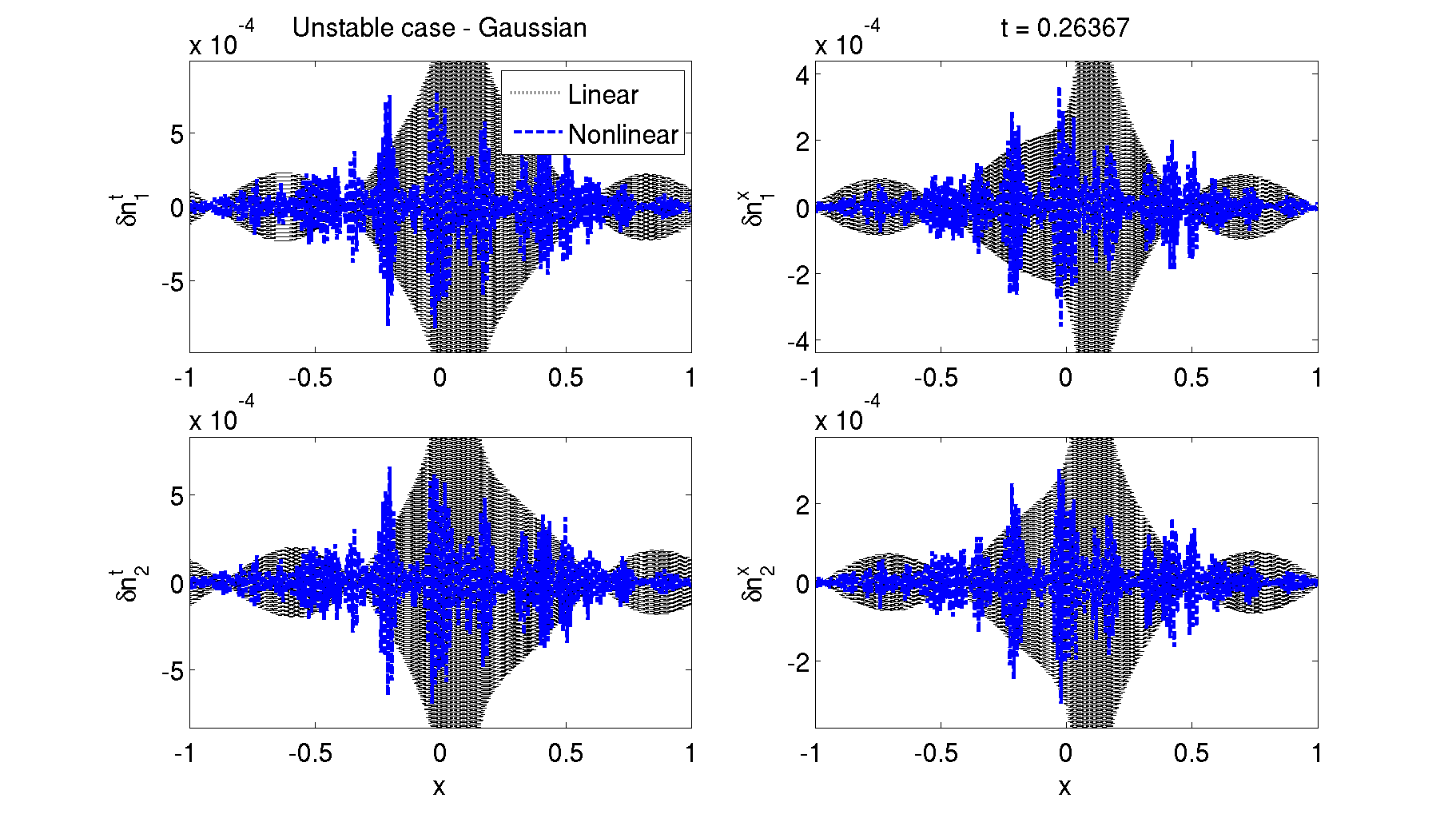}
  \caption{A direct comparison of the time evolution of the number density components in the coupled case within the instability window for the two-stream instability. The linearized solution is given by the dotted black lines, and the nonlinear numerical evolution (using 512 points - every point shown) is given by the blue dashed line. This is the entrainment case with $\kappa_{\Delta} = 0.5$ and a background velocity difference of $0.6$, as in Fig.~\ref{fig:Instability1}, but here the initial perturbation is a Gaussian, hence the higher resolution required. The instability grows rapidly in both linear and nonlinear case; to compare, the linear solution seeds all modes up to $k = 256$.}
  \label{fig:Instability1_Gaussian}
\end{figure}

We first consider the entrainment case with $\kappa_{\Delta} = 1/2$. As an illustration we look at a case where the first fluid is perturbed with a simple sine wave as before, but is at rest, whilst the second fluid is unperturbed but has a velocity of $0.6$ in the background so that the two-stream instability acts. Representative results are shown in Figs.~\ref{fig:Instability1} and \ref{fig:Instability1_Gaussian}. We note first that the space-time development of the instability appears to be (a) dominated by the high frequency components, (b) \emph{not} spatially localized to the perturbation (as shown in particular when the initial perturbation is a narrow Gaussian, as in Fig.~\ref{fig:Instability1_Gaussian}), and (c) closely following the linearized solution, where we assume that the linearized solution is ``seeded'' by constant frequency low amplitude perturbations in all modes in addition to the perturbation used in the numerical simulation.

\subsubsection{Time-frequency behaviour: linear case}
\label{sec:results_instability_linear}

\begin{figure}
  \centering
  \includegraphics[width=0.95\textwidth]{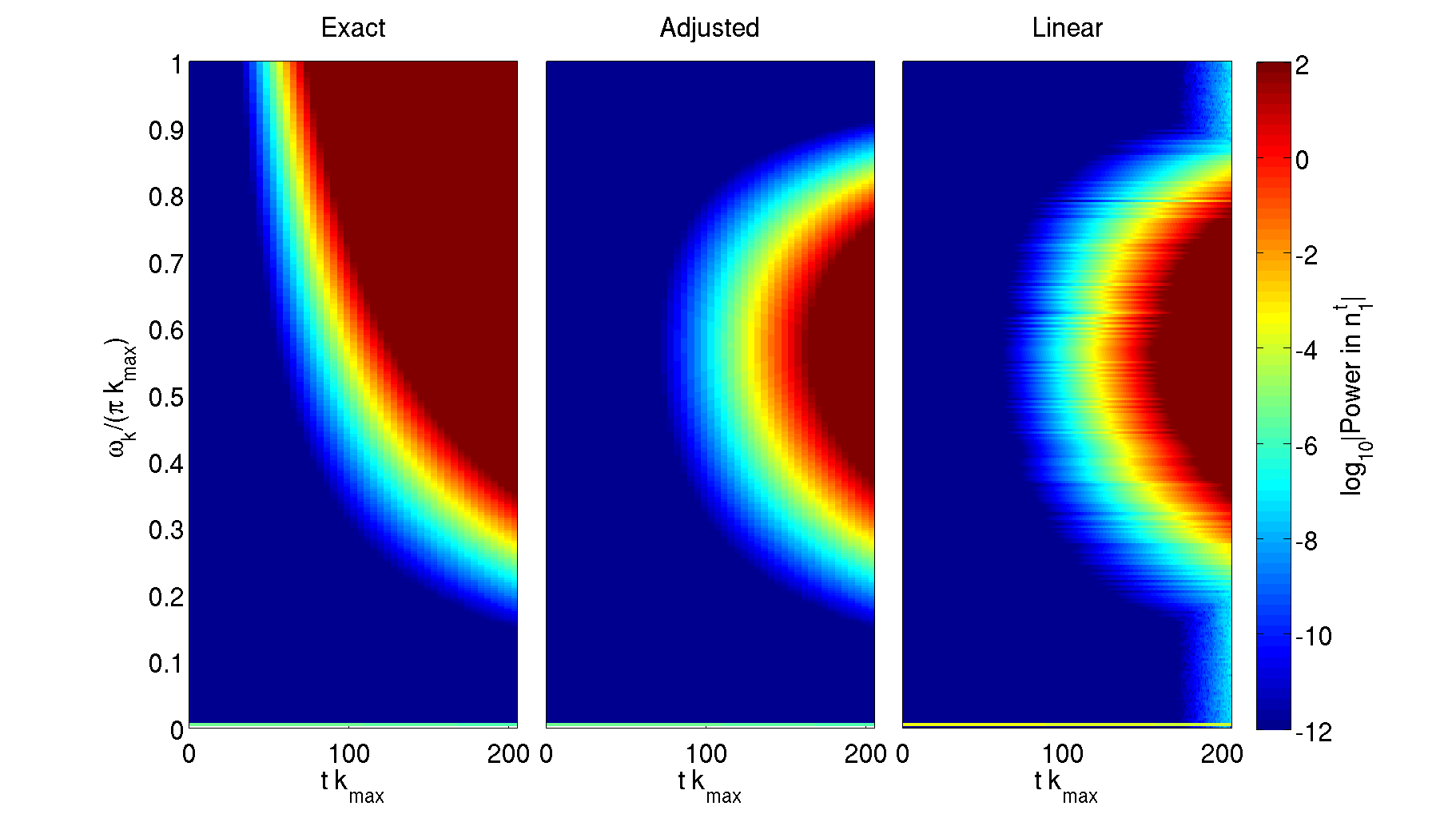}
  \caption{A time-frequency plot of the growth of the instability in the linearized case. Both time and frequency are scaled to be resolution independent -- the numerical simulation shown uses 1024 points ($k_{\text{max}} = 512$). The norm of the complex component of the discrete Fourier Transform of $\dnnu{1}{t}$ is shown -- the result for other components and quantities is qualitatively the same. The left plot shows the behaviour for the linearized solution. As expected the highest frequencies blow up the fastest. The central plot shows the behaviour for the linearized solution, adjusted as in Eq.~\eqref{eq:Fourier_numerical} to take into account the behaviour of the spatial differencing scheme. This would be the expected behaviour from a numerical evolution of the linearized equations. The right plot shows the results from the numerical \emph{linear} evolution. We see that the adjustment for the spatial differencing scheme completely explains the numerical behaviour, suppressing the growth of the instability at high frequencies.}
  \label{fig:Instability_FFT_Linear1}
\end{figure}

More detail can be seen when the solution is studied in time-frequency space.  These results, however, mix numerical and nonlinear effects. In order to disentangle these effects as far as possible we first consider solely the linearized case, as shown in Fig.~\ref{fig:Instability_FFT_Linear1}. In the linear case we expect the different frequency modes to behave independently. The change in power of a given frequency mode with time will therefore illustrate only effects due to the two-stream instability where relevant and the numerical method employed. The linearized solution, as constructed in Sec.~\ref{sec:linear_solution_solution}, is shown in the left panel and shows simple characteristic behaviour, with power-law behaviour in the instability growth time scale. This behaviour is not replicated in the numerical simulations of the \emph{linear} system, shown in the right panel. This must be purely due to the numerics used. This numerical effect can be modelled analytically.

Following the work of Lele~\cite{Lele1992} we note that the effect of the finite difference scheme, when approximating spatial derivatives, is equivalent to replacing the exact Fourier transform relation $\partial_x f \rightarrow \tj \omega \hat{f}$ with $\partial_x f \rightarrow \tj \omega' \hat{f}$, where $\omega'$ encodes how accurately the numerical method captures modes of a given frequency. For the centred fourth order differencing that we focus on here a straightforward calculation (or appropriately applying the general results of~\cite{Lele1992}) gives
\begin{equation}
  \label{eq:Fourier_numerical}
  \tilde{\omega}' (\tilde{\omega}) = \frac{8 \sin(\tilde{\omega}) - \sin(2 \tilde{\omega})}{6}.
\end{equation}
Here we use $\tilde{\omega} = \pi \omega / k_{\text{max}}$ to scale to Lele's units. Repeating the linear analysis using $\omega'$ in place of $\omega$ throughout leads to the results in the central panel of Fig.~\ref{fig:Instability_FFT_Linear1}. The match between these adjusted ``exact'' results and the results from numerical simulation in the right panel of Fig.~\ref{fig:Instability_FFT_Linear1} is extremely good. The remaining minor differences are likely due to the choice of using uniform amplitude ``noise'' to seed the analytic calculation at higher frequencies, as compared to the numerical simulation where the noise likely appears from floating point round-off error. This assumption is supported by the amplitude of the seed noise used in this comparison, which is $\approx 10^{-16} \times \delta$: the relative floating point accuracy of the linearized perturbation.

\subsubsection{Time-frequency behaviour: nonlinear case}
\label{sec:results_instability_nonlinear}

We next sketch our expectations of how the nonlinear effects would modify the behaviour. Considering a single fluid and taking solely the discrete Fourier Transform of the nonlinear system of Eq.~(\ref{eq:eom_3plus1}), we would expect the nonlinear modes to satisfy coupled equations of the \emph{qualitative} form
\begin{equation}
  \label{eq:nonlinear_solution_formal}
  \partial_t \tilde{n}^{a}_{[k]} + A_{[k]} \circledast \tilde{n}^{a}_{[k]} = 0,
\end{equation}
where $\circledast$ is the (cyclic) convolution product. In the linear case the matrix $A_{[k]}$ would only contain the constant (zero frequency, $k = 0$) term. In the nonlinear case the terms depend on the master function and the data -- in other words, on the non-trivial $\tilde{n}^{a}_{[k]}$.

\begin{figure}
  \centering  \includegraphics[width=0.95\textwidth]{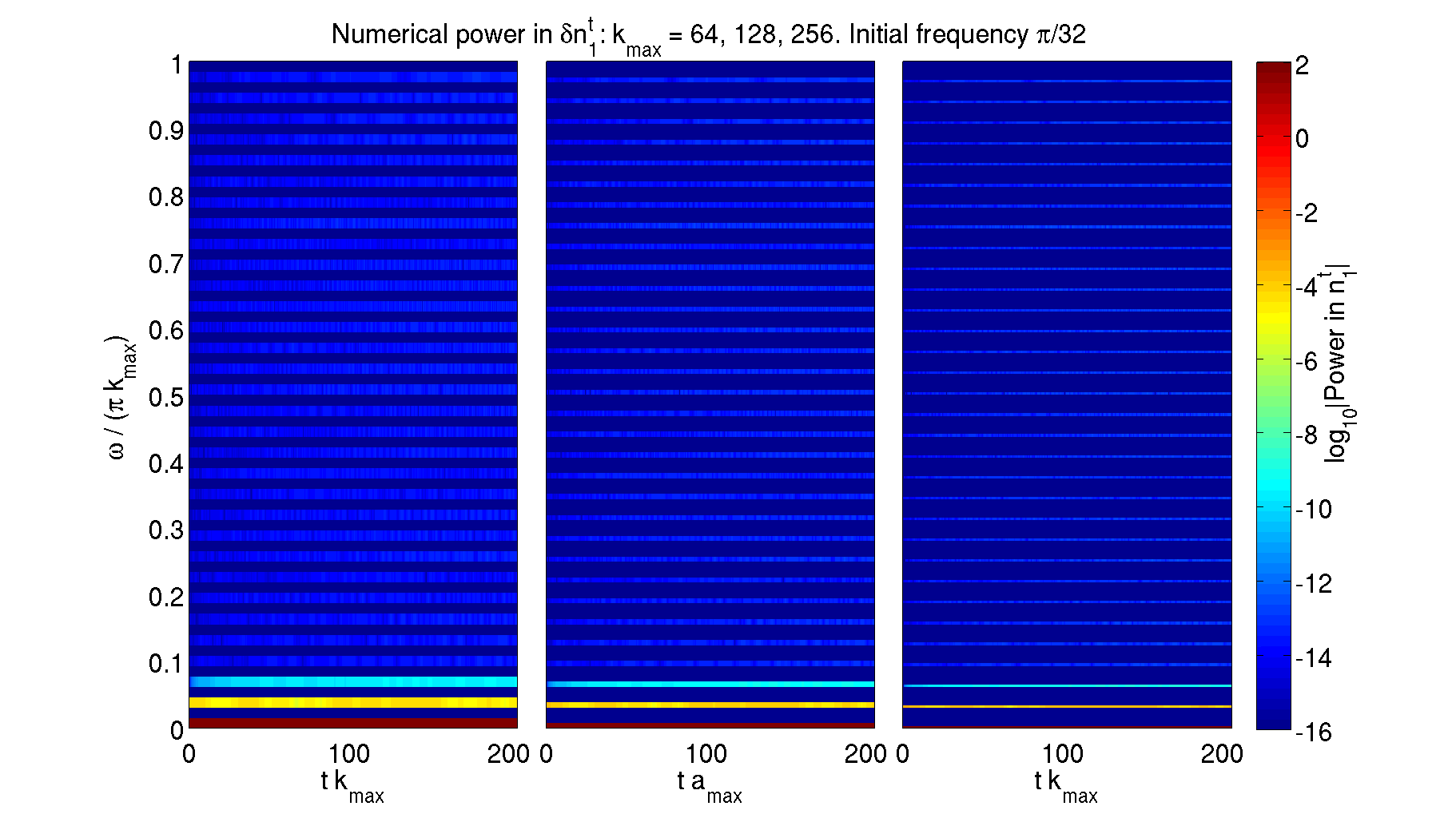}\\
  \includegraphics[width=0.95\textwidth]{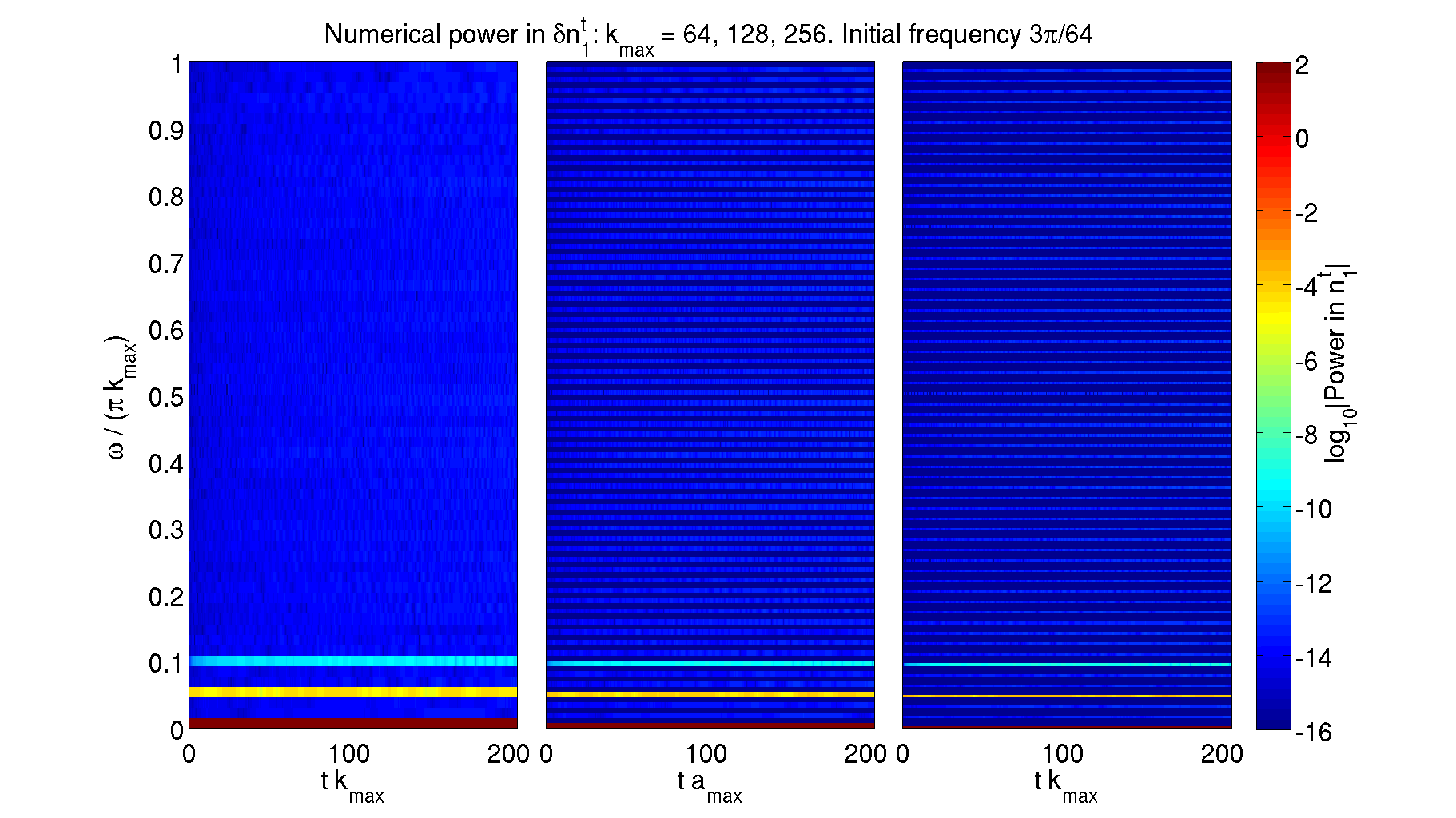}
  \caption{A time-frequency plot showing the harmonic coupling expected in the nonlinear case. As the grid resolution (and hence frequency maximum $k_{\text{max}}$) is varied the initial perturbation is modified to give a constant frequency with respect to the grid. In all cases the perturbation has a single frequency. The nonlinear coupling immediately excites all the higher harmonics, although only the lowest harmonic (at twice the frequency of the initial perturbation) is excited at a level that would be visible. The initial parameters are chosen such that the fluids couple but the two-stream instability is not acting.}
  \label{fig:NonlinearCoupling_FFT1}
\end{figure}

With initial data dominated by say two frequencies $k_1$ and $k_2$ the convolution couples modes at the harmonics with frequencies $a k_1 + b k_2$ with $a, b$ integers. For the typical initial data used here the dominant frequencies will be from the background ($k = 0$) and the initial perturbation with frequency $k_0$, leading to harmonics at integer multiples of $k_0$. This effect can be clearly seen in numerical nonlinear simulations as shown in Fig.~\ref{fig:NonlinearCoupling_FFT1}. In this case the background is chosen so that the velocity difference is $0.1$ so the two-stream instability does not act. The fluids couple and the higher harmonics are excited, but only the frequencies associated with the background, the initial perturbation and the first harmonic would be noticeable in the spatial snapshots.

\begin{figure}
  \centering
  \includegraphics[width=0.95\textwidth]{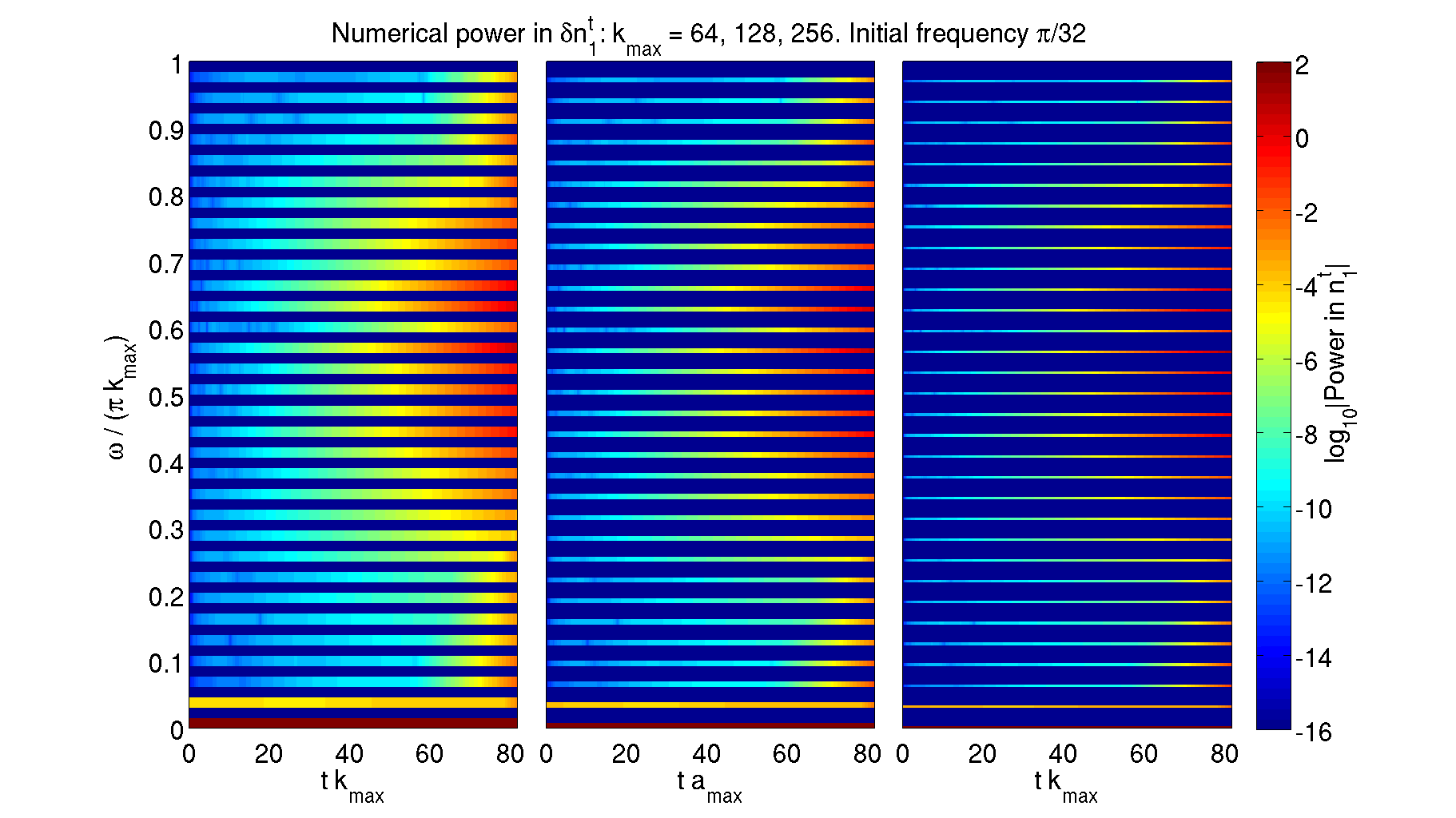}\\
  \includegraphics[width=0.95\textwidth]{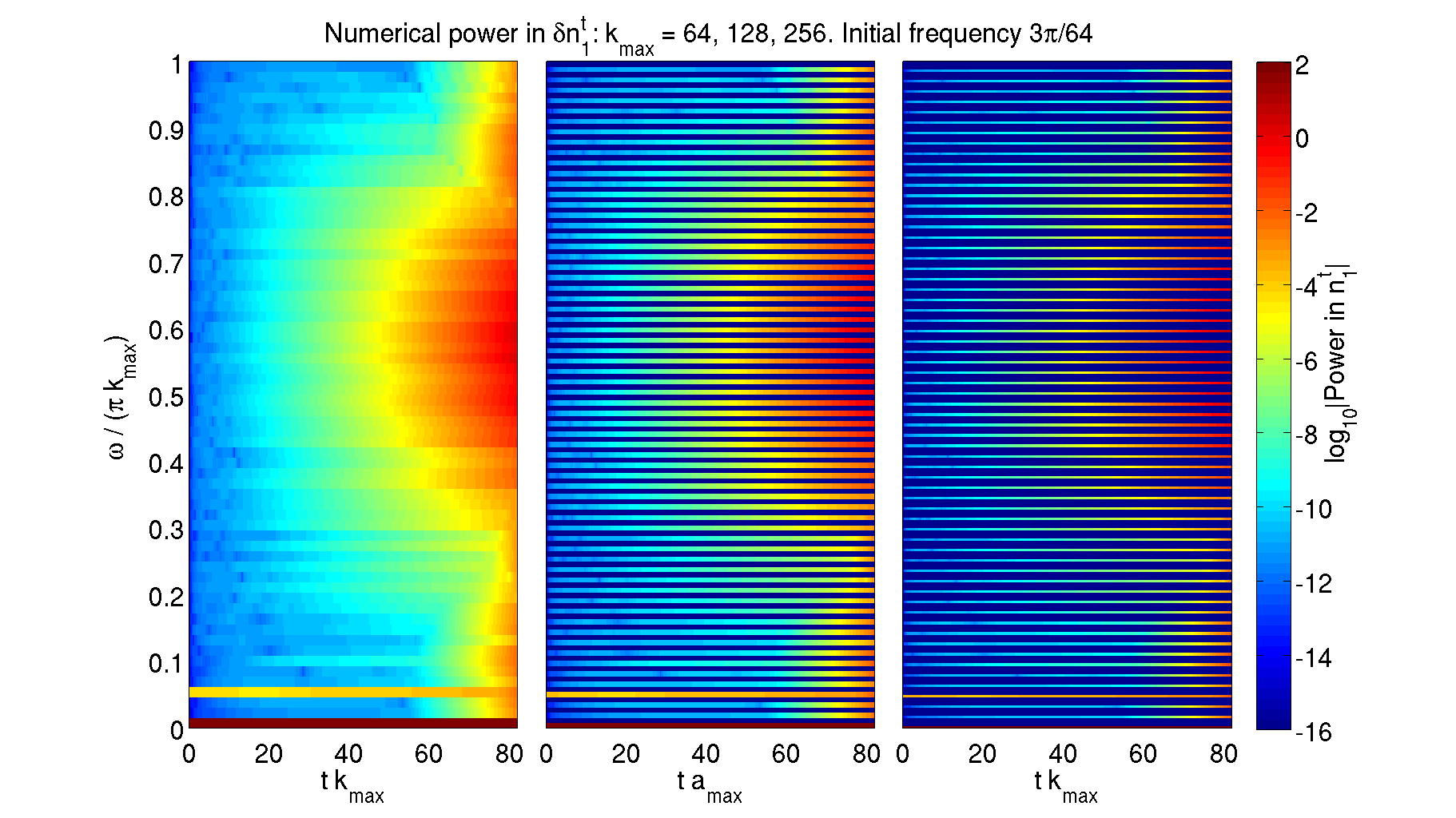}
  \caption{A time-frequency plot of the growth of the instability where the initial frequency (relative to the grid) of the perturbation is held fixed. As the grid size is increased the same number of harmonics are excited by the nonlinear coupling, as in the stable case shown in Fig.~\ref{fig:NonlinearCoupling_FFT1}. Each excited harmonic then grows exponentially due to the two-stream instability, with the pattern resembling (at early times) the linearized case shown in Fig.~\ref{fig:Instability_FFT_Linear1}). As in the stable nonlinear case in Fig.~\ref{fig:NonlinearCoupling_FFT1} there is no visible nonlinear coupling except to the immediate harmonics. In the top row the overtones are always at higher frequencies. The choice of period in the bottom row means that some overtones at lower frequencies are excited, meaning that in the bottom left plot \emph{all} frequencies on the grid are excited by nonlinear couplings.}
  \label{fig:Instability_FFT_2}
\end{figure}

Once the background is modified to allow the two-stream instability to act we can see both the numerical suppression of the high frequency modes (as shown in the linear case in Fig.~\ref{fig:Instability_FFT_Linear1}) and the nonlinear coupling of the higher harmonics (as shown in the stable case in Fig.~\ref{fig:NonlinearCoupling_FFT1}). Fig.~\ref{fig:Instability_FFT_2} repeats the calculation exciting higher harmonics as in Fig.~\ref{fig:Instability_FFT_Linear1}, but the background is modified so that the velocity difference is $0.6$ and the two-stream instability acts. Whilst all modes are excited to some degree by numerical error, only those that are coupled via harmonic overtones are excited to a level that is numerically significant (i.e.\ greater than $10^{-12}$ here). The figure then shows precisely these excited modes growing exponentially as expected, initially independently of the other modes, with the suppression at higher frequency qualitatively matching the behaviour seen in the linear case.

\begin{figure}
  \centering
  \includegraphics[width=0.95\textwidth]{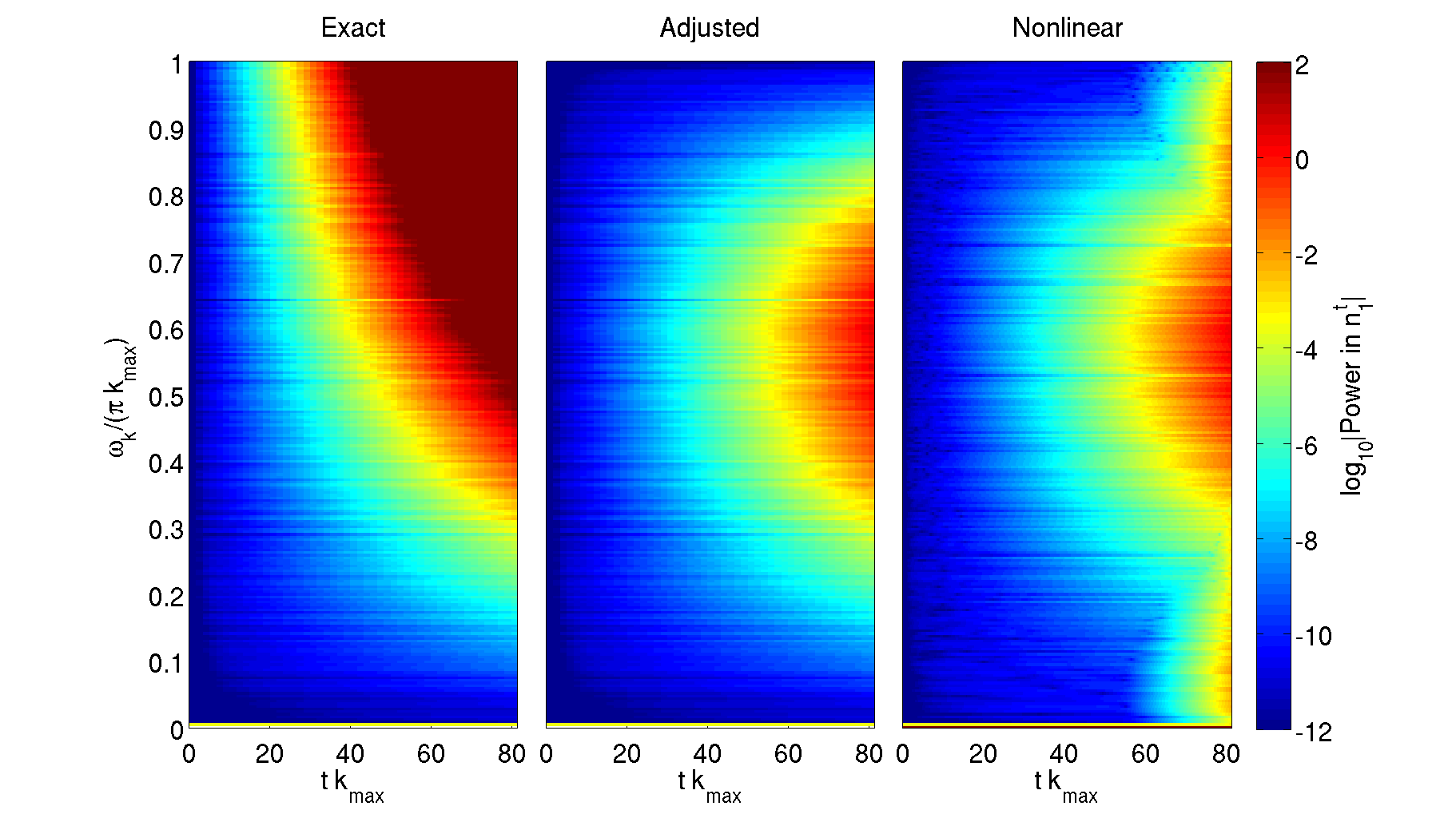}
  \caption{A time-frequency plot of the growth of the instability in the nonlinear case. This figure is the nonlinear equivalent of Fig.~\ref{fig:Instability_FFT_Linear1}. The linearized and adjusted solutions are seeded with noise from the first timestep of the numerical solution; this is noticeably larger than in the linear case. The qualitative comparison between the adjusted and numerical solution is still reasonable, explaining the suppression of the growth at high frequencies, but the nonlinear coupling does modify the solution significantly.}
  \label{fig:Instability_FFT_Nonlinear1}
\end{figure}

We can now look at the original case, studied for the linear problem in Fig.~\ref{fig:Instability_FFT_Linear1}. Here the only mode initially excited is the lowest mode on the grid. As seen in Fig.~\ref{fig:Instability_FFT_Nonlinear1}, there is still a qualitative match between the adjusted linear solution and the numerical evolution, even in the nonlinear case. The ``bulk'' features, including the approximate growth rates and the suppression of the growth at high frequencies, remain the same. Whilst there are also noticeable differences at late times, there is no clear pattern in either the space-time development or the time-frequency growth, except for the increased coupling between the different frequencies at late time.

\subsection{The instability onset}
\label{sec:results_instability_onset}

The parameters chosen for both the entrainment and chemical coupling case  have a ``window'' where the two-stream instability acts, as shown by Fig.~\ref{fig:background_stability}. By modifying the relative velocity of the background in the initial data, we can check that the onset of the instability occurs at the same point with the nonlinear code.

In both the entrainment and chemical coupling case we find that, when using small perturbations of amplitude $\delta = 10^{-6}$ as above, the onset of the instability (with increasing relative velocity) is, to numerical precision, identical to that predicted from the linear analysis. In the chemical coupling case we can also check the upper edge of the window; that is, the nonlinear results show the instability for $0.29185 \lesssim \Delta v \lesssim 0.69985$, and are stable otherwise. In the entrainment case we are unable to check the upper edge of the window (which is at $\Delta v \approx 0.958$) as the velocities required are too large for the nonlinear code to successfully evolve.

Finally we attempted to produce initial data that would generate shocks before the two-stream instability becomes important, to see whether the nonlinear couplings could ever be expected to be more important than the instability growth. In our experiments, only extremely large initial perturbations ($\delta \sim 10^{-1}$), combined with a choice of master function parameters such that the two-stream instability growth is as slow as possible, show signs of characteristic breaking before the two-stream instability sets in. Even in these cases it is possible for the two-stream instability to dominate simply by increasing the grid size, and so admitting grid frequencies that grow sufficiently rapidly. Within this purely ideal hydrodynamics case using the shearing box approximation it does not seem possible to suppress the instability.

\subsection{Higher dimensions}
\label{sec:results_2d}

All the results so far have been restricted to $1+1$ dimensions in the shearing box (periodic boundaries) approximation. We have performed numerical simulations that retain the shearing box approximation in higher dimensions -- in particular, detailed comparisons in $2+1$ dimensions were calculated. No qualitative differences were found. This can be understood by transforming to the frame of the background flow for one species. In this frame, there are only two possible effects from the additional spatial dimension. 

First, the angle between the perturbed flow and the background velocity difference will change the coupling. This is indeed the case: our simulations confirm that the projection of the perturbation onto the velocity difference must be non-zero for the instability to act. However, in this (generic) case, the behaviour of the instability growth is qualitatively identical to the $1+1$ dimensional case.

Second, the coupling between the species should potentially lead to a change in the alignment between the species, modifying the relative velocity between them. This could potentially change the growth of the instability. However, explicit simulations performed by modifying the angle of the perturbation with respect to the relative flow showed no visible difference in, for example, the growth of $\Delta^2$. Thus it does not seem possible for purely local hydrodynamic effects, even at the nonlinear level, to modify the behaviour of the instability.

\section{Discussion}
\label{sec:discussion}

The two-stream instability has been mooted as an explanation for a range of astrophysical applications from GRBs and pulsar glitches to cosmology. We have used numerical simulations to study the nonlinear development of this instability when the species are modelled using coupled relativistic hydrodynamics.

Our simulations show that the onset of the instability when only the local, purely hydrodynamic behaviour is considered, perfectly matches the predictions of linear theory. The restricted analysis in $1+1$ dimensions is sufficient, provided the initial data is appropriately interpreted, when there is no change in either the spacetime, the external forces, or the coupling parameters. When these restrictions are relaxed, for example in the cosmological case considered by~\cite{Comer2012} where the expansion of the universe is modelled by varying the coupling parameters, then the instability growth can be curtailed.

The initial growth of the instability is also well described by linear theory, after adjustment to account for the known properties of the numerical simulations. At the very late stages where the instability dominates is there a noticeable difference between the linear and nonlinear behaviour, which has no obvious pattern.

This work is a necessary first step towards generic nonlinear simulations of relativistic multifluids. As multifluid effects such as entrainment are expected to be important in, for example, heat conduction~\cite{Andersson2011}, charge conduction and resistivity~\cite{Andersson2012a}, and neutron superfluids~\cite{Andersson2012b}, high accuracy nonlinear simulations incorporating detailed microphysics must include multifluids. The main technical barrier to incorporating these effects in current simulations is the lack of either a balance law form or a non-conservative form that explicitly controls entropy at discontinuities. Our future work is aimed at overcoming this technical hurdle before combining the multifluid effects with nonlinear relativistic elasticity~\cite{Gundlach2012} and appropriate numerical techniques for interfaces~\cite{Millmore2010} to produce such simulations.

\bibliography{MultiFluid}

\end{document}